# Fundamentals and Applications of Time-varying Media: A Review

Youxiu Yu[1,†], Hao Hu[1,†,*], Qianru Yang[2], Linyang Zou[3], Dongjue Liu[4], Hao Chi Zhang[2], and Yu Luo[1,*]

[1]*National Key Laboratory of Microwave Photonics & College of Electronic and Information Engineering, Nanjing University of Aeronautics and Astronautics, Nanjing 211106, China*
[2]*State Key Laboratory of Millimeter Waves, Southeast University, Nanjing 210018, China*
[3]*School of Electrical and Electronic Engineering, Nanyang Technological University, Singapore 639798, Singapore*
[4]*National Key Laboratory of Scattering and Radiation, Shanghai Radio Equipment Research Institute, Shanghai 200438, China*

[†]*These authors contribute equally*
*Corresponding author: hao.hu@nuaa.edu.cn (Hao Hu) and yu.luo@nuaa.edu.cn (Yu Luo)*

**ABSTRACT**

Time-varying media, characterized by dynamic or spacetime-modulated constitutive parameters such as permittivity and permeability, have recently emerged as a transformative paradigm for advanced wave control, transcending the constraints imposed by temporal translation symmetry and energy conservation in static systems. By incorporating time as an active degree of freedom, such media unlock unique phenomena including broadband frequency conversion, temporal refraction, significant field enhancement, and magnet-free nonreciprocity. These capabilities are reshaping the landscape of photonic technologies, enabling groundbreaking applications such as broadband nonreciprocal amplifiers, non-resonant lasers, and highly efficient particle accelerators. This review systematically classifies time-varying media based on their modulation schemes and elucidates the underlying physical principles and distinctive wave-matter interactions. We comprehensively survey the latest advances in this rapidly evolving field, highlighting exotic wave behaviors and practical implementations across electromagnetic and photonic systems. Furthermore, we summarize experimental platforms that realize time-varying responses across different frequency regimes. Finally, we assess the current state of progress, identify key challenges, and offer a forward-looking perspective on future research directions in this dynamic and promising area.

## I. INTRODUCTION

The study and application of electromagnetic waves have long mirrored the course of scientific and technological progress.[1] As a channel for energy storage and transfer, the fundamental nature of electromagnetic waves was revealed through Maxwell's theoretical framework and Hertz's experimental validation.[2] This foundational understanding has provided the theoretical basis for the applications of electromagnetics. With the growing demand for precise control of electromagnetic waves, material properties have become central to enable diverse wave-manipulation functionalities. However, natural existing materials typically exhibit finite constitutive parameters, such as conductivity, permittivity and permeability, making them challenging to fulfill the diverse needs for wave manipulation.

The emergence of artificially engineered media significantly expanded the electromagnetic material functionalities, making their development a central focus of modern research.[3,4] A representative example is the proposal of metamaterials in the 1996, which are generally realized by structuring matters on subwavelength scales.[5,6] The advent of metamaterials not only provides a feasible solution to achieve negative-index materials that are proposed in 1960s,[7–9] but also gives rise to many unconventional effects such as super-resolution imaging,[10] high-efficiency absorption,[11] and enhanced spontaneous or stimulated light emission, etc.[12] Later in 2006, the foundation of transformation optics offers a universal scheme to guide the design of metamaterials, enabling the redirection of electromagnetic waves at will.[13,14] Since then, the invisibility cloaks become a major research direction and application in the community of metamaterials.[15] Equally important, photonic crystals, first proposed in the 1980s, provide an alternative pathway for controlling electromagnetic waves.[16,17] The photonic crystals are materials arranged periodically in space and its dispersion is generally characterized by a frequency bandgap, within which wave propagation is prohibited.[18,19] This unique property has been widely exploited in applications such as photonics crystal waveguiding, spontaneous emission control, and sensing.[20–22] Moreover, since 2008 when quantum Hall effect has been transferred from quantum to classical systems,[23,24] the frequency bandgap in photonic crystals is found to be associated with a quantized topological invariant, provided the system preserves certain symmetries (e.g., time-reversal symmetry, particle-hole symmetry, and chiral symmetry).[25] Such systems are classified as photonic topological insulators, where the topological invariants serve as predictors for the emergence of diverse topological states (e.g., edge states, corner states, and hinge states) under open boundary conditions.[26] Owing to symmetry protection, these states exhibit remarkable robustness against disorders and defects, enabling potential applications in topological directional coupling, nonreciprocal light transport, lasing, and quantum information processing.[27–30] Despite these remarkable achievements, the manipulation of electromagnetic waves in the above artificially engineered materials often suffers from limited reconfigurability and environmental adaptability due to time-invariant nature.

Driven by the urgent demand for dynamic wave manipulation, time-varying media have recently been recognized as a vital member in the big family of artificially engineered materials.[31–35] Time-varying media are defined as ones whose constitutive parameters are abruptly or continuously changed in time. Early in the 1950s, time was considered as a novel degree of freedom for controlling waves.[36–38] However, due to the lack of technologies capable of rapidly and controllably altering the electromagnetic properties at the macroscopic scale, early research on this topic remained stagnant. Until 2010s, time-varying media experienced rapid development thanks to the maturity of technologies like high-speed switches and programmable metamaterials.[31,39,40] This work primarily focuses on the propagation and scattering behaviors of electromagnetic waves in bulk material systems under time modulation and spacetime modulation. In this context, moving interfaces are formed through dynamic modulation of bulk media, and specific experimental implementation schemes are discussed in Section IV. Active metasurfaces exhibit significant potential for wavefront manipulation and system integration and have been thoroughly reviewed;[41,42] here, they are mentioned only in connection with the relevant phenomena and their possible implementation via metasurfaces. Spacetime interfaces form a key basis of time-varying media. Generally, as illustrated in Fig. 1(a), spacetime interfaces with a constant interfacial velocity could be classified into three types: subluminal ($|\beta| < \min\{1/n_i\}$), interluminal ($\min\{1/n_i\} \leq |\beta| \leq \max\{1/n_i\}$), and superluminal ($|\beta| > \max\{1/n_i\}$) ones, where $\beta = v/c$ is the interface velocity normalized to the light speed in vacuum, and $n_i$ are the refractive indices of surrounding media.[39] In the extreme case of $\beta = 0$, where the time modulation vanishes, the subluminal interface degenerates into a spatial interface; conversely, if the modulation occurs exclusively in the temporal dimension (corresponding to $\beta = \infty$), the superluminal interface becomes temporal interface. Here, subluminal, interluminal, and superluminal interfaces all belong to spacetime modulation, whereas spatial and temporal interfaces correspond to space and time modulations, respectively. It should be emphasized that time modulated media is a special subclass of time-varying media, as its constitutive parameters are entirely dependent on time. Spacetime modulated media are characterized by constitutive parameters that evolve in both space and time. Among all the time-varying media, the periodically modulated system is of particular interest due to the enhanced degree of freedom for dispersion engineering. The modulation function of a constitutive parameter of a periodically modulated system could be generally expressed as $\cos(K_i x - \Omega_i t) + \Gamma \cos(K_i x + \Omega_i t)$, where $K_i$, $\Omega_i$ and $\Gamma$ correspond to the modulated wavevector, modulated frequency, and proportionality factor, respectively. We denote such a modulation type as traveling-standing-wave modulation, with the standing wave radio $\text{SWR} = |\frac{1+\Gamma}{1-\Gamma}|$. The SWR=1, if $\Gamma = 0$ or $\Gamma = \infty$, corresponding to the traveling-wave modulation, as shown in the Fig. 1(b). The modulation function of such a system takes the form of $\cos(K_i x - \Omega_i t)$ or $\cos(K_i x + \Omega_i t)$. On the other hand, SWR=∞, if $\Gamma = 1$, corresponding to the standing-wave modulation. Such a system is featured as a modulation function of $\cos(K_i x)\cos(\Omega_i t)$. Except for the above classifications, another major group belongs to random modulation, which has been almost unexplored in time-varying media so far [Fig. 1(c)]. The diverse strategies of modulation not only extend material concepts such as photonic crystals and metamaterials from space to the spacetime domain, but also overcome the conventional physical limits imposed by temporal translation symmetry, thereby enabling distinctive phenomena such as broadband frequency conversion, time reversal, magnet-free nonreciprocity, and parametric amplification.[31,43–49]

This review provides a comprehensive overview of the recent developments in the field of time-varying media, highlighting their significance in electromagnetic systems. We first present a systematic analysis of their classifications, fundamental physical principles, and properties. Subsequently, we outline the phenomena and applications of the time-varying media. Additionally, we summarize experimental strategies to implement time-varying media in different frequencies. Finally, we present the key developments and persistent challenges in this field, and propose potential directions for future investigations.

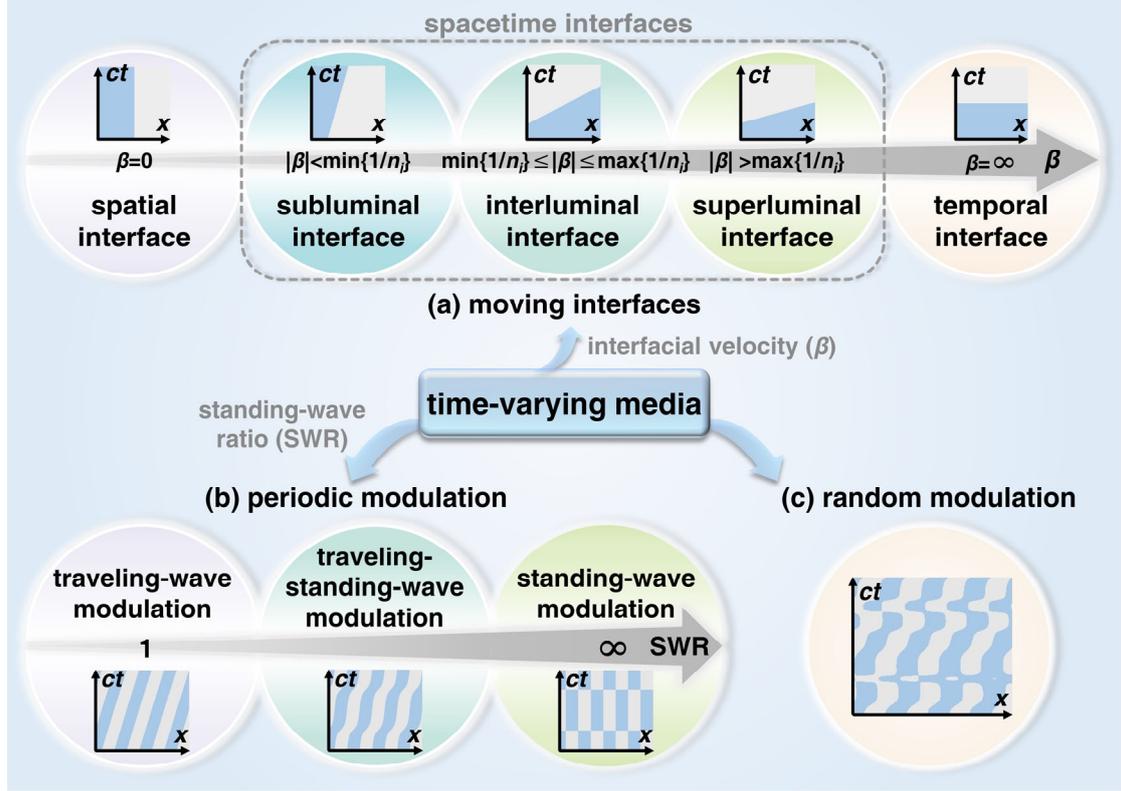

**FIG. 1.** Schematic diagram of time-varying media classification. (a) Classification of moving interfaces according to interfacial velocity. The moving interface is classified into three types: subluminal interface ($|\beta|<\min\{1/n_i\}$), interluminal interface ($\min\{1/n_i\}\leq|\beta|\leq\max\{1/n_i\}$), and superluminal interface ($|\beta|>\max\{1/n_i\}$), where $\beta$ is the ratio of interface velocity $v$ and light velocity in vacuum $c$, and $n_i$ is the refractive index of surrounding medium. The subluminal interface transforms into a spatial interface when $\beta\rightarrow 0$; if $\beta\rightarrow\infty$, the superluminal interface becomes a temporal interface. (b) Classification of periodic modulation according to standing-wave ratio. The modulation function of a constitutive parameter is characterized by $\cos(K_i x-\Omega_i t)+\Gamma\cos(K_i x+\Omega_i t)$, where $K_i$, $\Omega_i$ and $\Gamma$ represent modulated wavevector, modulated frequency and proportionality factor, respectively. Traveling-standing-wave modulation is defined by the standing wave ratio ($\text{SWR}=|\frac{1+\Gamma}{1-\Gamma}|$). SWR=1 corresponds to the traveling-wave modulation, while $\text{SWR}\rightarrow\infty$ indicates the standing-wave modulation. (c) Schematic of random modulation.

## II. PRINCIPLES OF SPACETIME INTERFACES AND MODULATIONS

To begin with, we reveal the influence of modulation schemes on the scattering behaviors of electromagnetic waves in systems with uniform-velocity moving interfaces. Without loss of generality, we consider a system with the spatial (i.e., stationary) interface, temporal interface, and spacetime interface, respectively, on the two sides of which the corresponding homogeneous medium refractive indices are denoted as $n_1$ and $n_2$, respectively. Generally, during the interaction between a wave and a stationary interface, the reflected and transmitted waves appear in different media, namely, on the both sides of this interface [Fig. 2(a)]. This scattering behavior remains the same in the system of subluminal interface [Fig. 2(c)]. In stark contrast, temporal and superluminal interfaces exhibit unique scattering phenomena that the reflected and transmitted waves coexist within the same medium due to the constraints of causality [Figs. 2(b) and 2(d)]. Despite similar scattering behaviors at the stationary (temporal) and subluminal (superluminal) interfaces, the corresponding frequency and wavevector conversions exhibit significant differences. As shown on the right of Fig. 2(a) [Fig. 2(b)], in systems with space (time) modulations, the frequency (wavevector) is conserved owing to the preserved temporal

(spatial) translation symmetry, whereas the wavevector (frequency) is not conserved, i.e., $\Delta\omega = 0$, $\Delta k \neq 0$ ($\Delta\omega \neq 0$, $\Delta k = 0$). However, neither frequency nor wavevector is conserved in the system with spacetime modulations [the right panels of Figs. 2(c) and 2(d)], since both temporal and spatial translation symmetries are simultaneously broken. Specifically, under spacetime modulations, the frequencies and wavevectors of scattered waves from subluminal and superluminal interfaces are determined by the phase-matching conditions $\Delta k v - \Delta \omega = 0$.

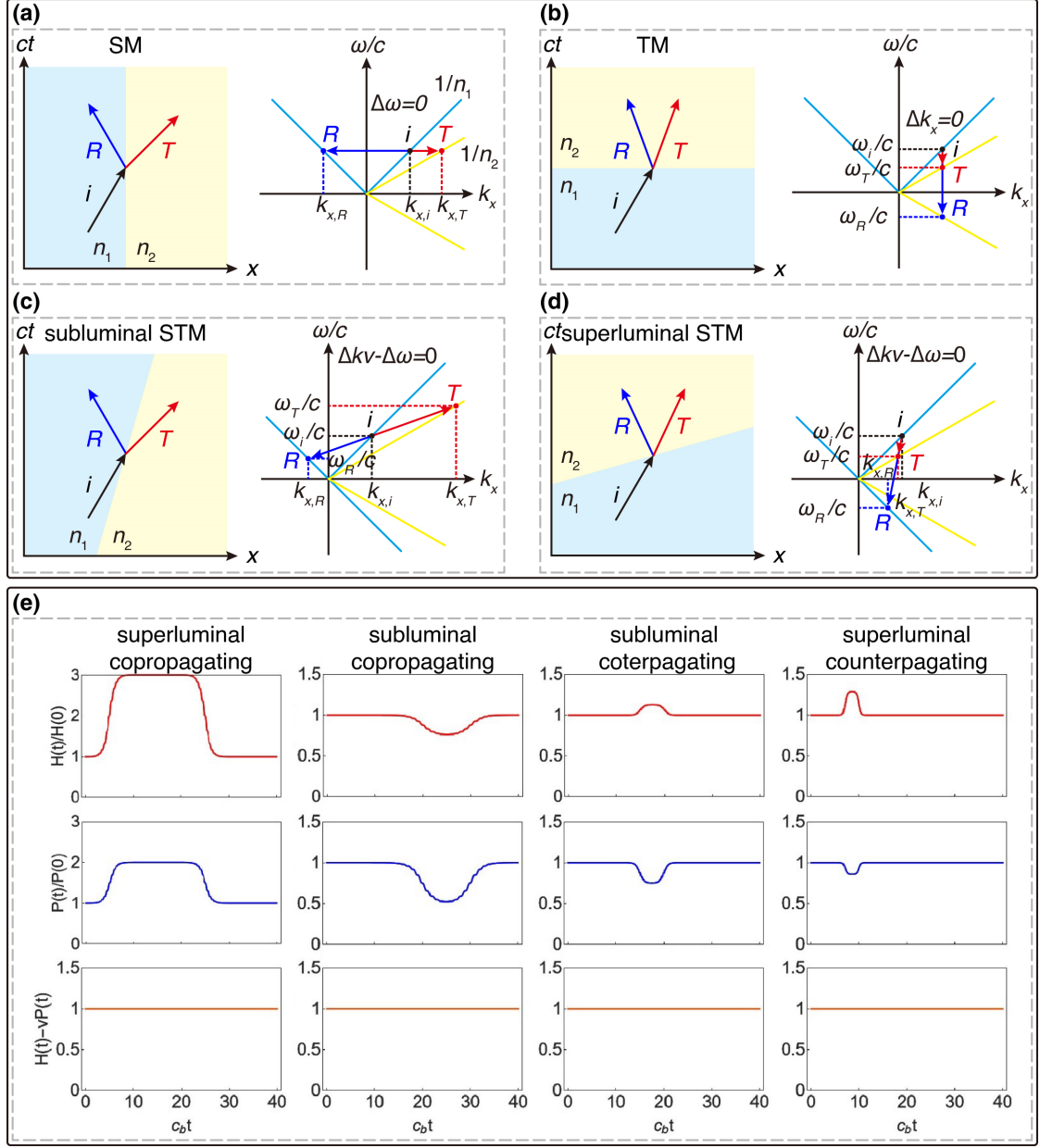

**FIG. 2.** The principles of time-varying media. Scattering phenomena and frequency conversion at a single interface under space modulations (a), time modulations (b), subluminal uniform-velocity spacetime modulations (c), and superluminal uniform-velocity spacetime modulations (d). (e) Conserved quantities under uniform-velocity spacetime modulations. Reproduced with permission from Liberal et al., ACS Photonics **11**(12), 5273-5280 (2024). Copyright 2024 Authors; licensed under a Creative Commons Attribution (CC BY) license.[50]

The amplitudes of these scattered waves are determined by the spacetime boundary conditions $\Delta \boldsymbol{E} + \boldsymbol{v} \times \Delta \boldsymbol{B} = 0$ and $\Delta \boldsymbol{H} - \boldsymbol{v} \times \Delta \boldsymbol{D} = 0$, where $\boldsymbol{E}$, $\boldsymbol{H}$, $\boldsymbol{D}$, and $\boldsymbol{B}$ are the electric field, magnetic field, electric displacement, and magnetic flux density, respectively. Under space modulations, the boundary conditions reduce to $\Delta \boldsymbol{E} = 0$ and $\Delta \boldsymbol{H} = 0$, whereas under time modulations, they become $\Delta \boldsymbol{B} = 0$ and $\Delta \boldsymbol{D} = 0$. The frequencies, wavevectors, and scattering coefficients of waves, due to the interaction with the dynamic interface, are summarized in Table 1.[51,52] As illustrated in columns 2 and 3 of Table 1, under space and time modulations, the frequencies, wavevectors, and scattering coefficients solely depend on the refractive index. Except for the refractive index, the interface velocity is also a key parameter influencing the frequencies, wavevectors, and scattering coefficients in spacetime modulations [columns 4 and 5 of Table 1].

**Table 1.** The summary of frequencies, wavevectors, and scattering coefficients from a single interface under different modulation types.

| coefficients | space modulations | time modulations | uniform-velocity spacetime modulations | |
|---|---|---|---|---|
| | | | subluminal | superluminal |
| frequency | $\omega_r = \omega_t = \omega_i$ | $\omega_r = -\dfrac{n_1}{n_2}\omega_i$ | $\omega_r = \dfrac{1-n_1\beta}{1+n_1\beta}\omega_i$ | $\omega_r = \dfrac{1-n_1\beta}{1+n_2\beta}\omega_i$ |
| | | $\omega_t = \dfrac{n_1}{n_2}\omega_i$ | $\omega_t = \dfrac{1-n_1\beta}{1-n_2\beta}\omega_i$ | $\omega_t = \dfrac{1-n_1\beta}{1-n_2\beta}\omega_i$ |
| wavevector | $k_{z,r} = -k_{z,i}$ | $k_{z,r} = k_{z,t} = k_{z,i}$ | $k_{z,r} = -\dfrac{1-n_1\beta}{1+n_1\beta}k_{z,i}$ | $k_{z,r} = \dfrac{1-n_1\beta}{1+n_2\beta}k_{z,i}$ |
| | $k_{z,t} = \dfrac{n_2}{n_1}k_{z,i}$ | | $k_{z,t} = \dfrac{n_2}{n_1}\dfrac{1-n_1\beta}{1-n_2\beta}k_{z,i}$ | $k_{z,t} = \dfrac{n_2}{n_1}\dfrac{1-n_1\beta}{1-n_2\beta}k_{z,i}$ |
| reflection | $R = \dfrac{n_2-n_1}{n_1+n_2}$ | $R = \dfrac{n_2-n_1}{2n_2}$ | $R = \dfrac{n_2-n_1}{n_1+n_2}\dfrac{1-n_1\beta}{1+n_1\beta}$ | $R = \dfrac{n_1-n_2}{2n_1}\dfrac{1-n_1\beta}{1+n_2\beta}$ |
| transmission | $T = \dfrac{2n_2}{n_1+n_2}$ | $T = \dfrac{n_1+n_2}{2n_2}$ | $T = \dfrac{2n_2}{n_1+n_2}\dfrac{1-n_1\beta}{1-n_2\beta}$ | $T = \dfrac{n_1+n_2}{2n_1}\dfrac{1-n_1\beta}{1-n_2\beta}$ |

Among the scattering behaviors at a variety of dynamic interfaces, the one in the interluminal regime is particularly distinctive, as identified in 1975.[53] Here, we analyze the scattering behavior of waves incident from a low- to high-refractive index medium ($n_1 \to n_2$, $n_1 < n_2$) for example. As illustrated in Fig. 3(a), only a reflection could be induced when the incident wave co-propagates with the interface. The underlying reason is that the transmitted wave propagates too slowly in the high-refractive index medium ($v_2 < |\beta|c$) such that it is suddenly trapped by the interface once generated. On the other hand, three scattered waves (i.e., two reflected and one transmitted) arise when the incident wave counter-propagates with the interface [Fig. 3(b)]. The coexistence of twin reflections is physically permitted because the interface moves slower than the reflection in the incident medium but faster than the reflection in the transmitted medium ($v_2 < |\beta|c < v_1$). Furthermore, a single reflection or a single transmission occurs when the incident wave propagates from a high- to low-refractive index medium. However, this research direction faces a core bottleneck due to the theoretical debate and is therefore worthy of further exploration (see more discussions in the section of conclusion and outlook).

Next, we analyze the energy-momentum correlation in the system with uniform-velocity spacetime modulation. The phase-matching conditions require that $kv - \omega$ is a conserved quantity, which indicates that the conserved quantity in the system of uniform-velocity spacetime modulation is not an individual energy or momentum. Through the rigorous derivation of Noether's theorem, this conserved quantity corresponds to the difference between electromagnetic energy and interface velocity-weighted Minkowski momentum (i.e., $H(t) - vP(t)$), when the permittivity and permeability satisfy

$(\partial_t + v\partial_z)\varepsilon(z,t) = 0$ and $(\partial_t + v\partial_z)\mu(z,t) = 0$ .[50] This conserved quantity arises from the exact cancellation of energy/momentum sources and sinks induced by uniform-velocity spacetime modulation. As demonstrated in Fig. 2(e), the energy-momentum correlation holds universally under uniform-velocity spacetime modulation, irrespective of the modulation regimes (subluminal or superluminal) or the direction of interface motion. This phenomenon is markedly different from space modulations or time modulations.[54] In the system with space modulations, the energy is conserved, whereas the momentum is changed. On the other hand, in the system with time modulations, the momentum is invariant while the energy varies.

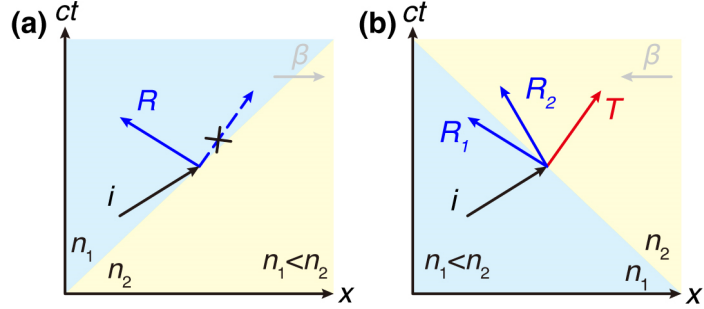

FIG. 3. Schematic diagram of scattering behavior in interluminal regime. (a) The scattering behavior when the interface moves in the same direction as the incident wave. (b) The scattering behavior when the interface moves in the opposite direction as the incident wave. Here, the refractive indices of the medium on two sides of interface are $n_1$ and $n_2$ ( $n_2 > n_1$ ), respectively.

To further understand the propagation characteristics of electromagnetic waves in systems with spacetime modulation, we compare two representative classes of dynamic systems, i.e., moving interfaces and moving matters. In moving interface systems, the interface moves relative to the observer while atoms and molecules merely oscillate around their equilibrium positions without net displacement relative to observers.[55] The moving interface is realized via external signal perturbations (e.g., pump pulses), and the effective interface velocity is angle-dependent. As a key merit, the moving interface systems enable superluminal modulation, which is initiated by an obliquely incident perturbation at the interface. This does not violate the fundamental rule that no form of energy can propagate faster than the speed of light in a vacuum, because the moving interface does not carry energy. In moving matter systems, the atoms and molecules in the medium move relative to the observer but the interface is stationary. Unlike moving interface systems, the speed of moving matter is restricted by the light speed in vacuum since they possess energy.

### III. DESIGN AND APPLICATIONS OF TIME-VARYING MEDIA

To provide a comprehensive and systematic overview, studies on time-varying media can be broadly categorized into two major modulation paradigms, namely, time modulated media and spacetime modulated media. Building upon this classification, the following sections review the design and applications of time-varying media. Owing to the intrinsic complexity of spacetime modulation in both physical mechanisms and implementation strategies, this class is further divided into three representative research directions to review: spacetime interfaces and slabs, spacetime metamaterials, and spacetime photonic crystals.

#### A. Time modulated media

Time modulations have emerged as a novel platform for controlling electromagnetic waves, offering dynamic functionalities beyond those achievable in conventional space modulations. A wide range of classical electromagnetic concepts, such as double-slit diffraction,[56] antireflection coatings,[57] Brewster angle,[58] Faraday rotation,[48] effective medium theory,[59–61] Anderson localization,[62,63] harmonic generation,[64] topological phenomena[43,65–67] and Smith-Purcell radiation,[68] have been successively transferred into the time domain. The extension from space to time has also led to a variety of applications of time modulated media, including time aiming,[69] parametric amplification,[70] holography,[46] inverse prisms,[71] amplified emission,[72,73] and spontaneous emission.[74]

Systems made of one or more temporal interfaces are one of the most fundamental time modulated media. A particularly illustrative case of such a system is the temporal double-slit.[56] The temporal double-slit is constructed by 4 temporal interfaces and the refractive index of the host medium undergoes abrupt temporal changes twice. As shown at the bottom of Fig. 4(a), the interaction between the temporal double-slit and incidence leads to the temporal double-slit diffraction. Such a diffraction spectrally is reflected as the emergence of alternating bright and dark fringes in the frequency domain. This is highly different from the traditional Young's double-slit diffraction, where the fringe patterns appear only in the wavevector domain. Remarkably, the periodicity and visibility of these frequency-domain interference patterns are highly dependent on the sizes and profiles of the time slits, providing an accurate measure of the material's temporal response. Judiciously engineering the temporal interface also enables the elimination of time reflection. Time reflection occurs when an incident wave hits a temporal interface. Inspired from the principle of quarter-wave impedance transformer, its temporal analogue can be conceptualized as a temporal structure comprising two time interfaces separated by a specific time interval [Fig. 4(b)].[57] At specific refractive index and time duration of the intermediate temporal slab, the time reflections induced from two temporal interfaces undergo destructive interference, thus the overall time reflection disappears. This concept of temporal quarter-wave impedance transformer is applicable not only to homogeneous media but also to waveguides with distinct cross sections, facilitating seamless energy transfer in time-varying photonic circuits.

Moreover, leveraging the characteristic of time modulated media and conventional anisotropic media enable dynamic control and deflection of electromagnetic wave propagation.[69] As illustrated in Fig. 4(c), the transmitted wave can be steered toward a designated receiver in real time by dynamically switching the media from isotropic to anisotropic at a specific moment. Furthermore, this modulation strategy can effectively manipulate spin-orbit interactions and radiation from stationary charge, which provides a novel platform for the development of tunable devices.[75–77]

The advancement of time modulations has spurred exponential growth of interest in photonic time crystals (PTCs), which are realized through the periodic time modulation of constitutive parameters.[31,32] Owing to their periodic nature, PTCs show unique capability for dispersion engineering and energy manipulation, making it more flexible for wave control in time as compared to a single temporal interface. The bandgap in PTCs is opened in momentum or wavevector space in the dispersions, which is referred as momentum bandgap. This is different from spatial counterparts, where bandgap is opened in energy or frequency. These momentum bandgaps in PTCs enable field amplifications, while the energy bandgaps in conventional photonic crystals forbid wave transmission. The unique property of PTCs provides an indispensable platform to control emissions from dipoles or free electrons. For instance, a dipole embedded in a PTC can emit radiation wherein modes within the momentum bandgap undergo broadband exponential amplification [Fig. 4(d)].[72] Moreover, the decay rate of spontaneous emission could be significantly enhanced within the momentum bandgap.[74] Follow-up work has shown that the momentum bandgap can be drastically expanded to be semi-infinite via introducing time modulation in resonant materials [Fig. 4(e)], which exhibits either an intrinsic material resonance or a spatially structural resonance.[78] More surprisingly, the momentum bandgap could even cover the entire momentum-space by exploiting longitudinal plasmons in epsilon-near-zero (ENZ) materials.[79,80] Since the resonant frequencies of longitudinal plasmons are wavevector-independent, coupling two branches of longitudinal plasmons with suitable time modulation leads to the infinite momentum bandgap. Such an infinite momentum bandgap enables the synchronous optical amplification in the full-wave band. Furthermore, the temporal Fabry-Pérot resonance in PTCs can overcome the long-wavelength limitation of the conventional Maxwell-Garnett effective medium theory.[60] Under specific discrete spectral conditions, even when the temporal period of the incident wave is comparable to or shorter than the modulation period of the PTC, impedance-mismatched structures can still be effectively described as a homogeneous time slab [Fig. 4(f)].

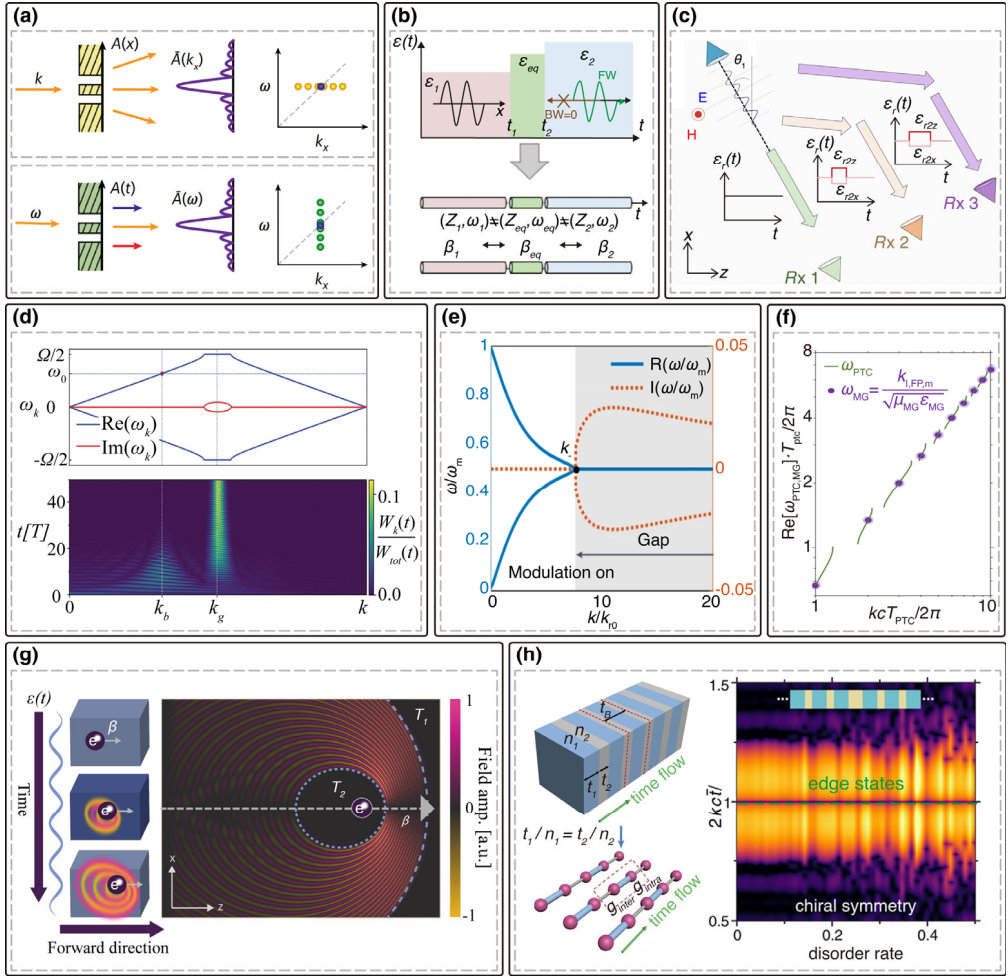

**FIG. 4.** Electromagnetic wave manipulation in time modulations. (a) Spatial (top) and temporal (bottom) double-slit diffraction. Reproduced with permission from Tirole et al., Nat. Phys. **19** (7), 999-1002 (2023). Copyright 2023 Authors; licensed under a Creative Commons Attribution (CC BY) license.[56] (b) Antireflection temporal coating. Reproduced with permission from Pacheco-Peña et al., Optica **7** (4), 323 (2020). Copyright 2020 Authors; licensed under a Creative Commons Attribution (CC BY) license.[57] (c) Temporal aiming. Reproduced with permission from Pacheco-Peña et al., Light Sci. Appl. **9** (1), 129 (2020). Copyright 2020 Authors; licensed under a Creative Commons Attribution (CC BY) license.[69] (d) The band structure and dipole emission in PTCs. Reproduced with permission from Lyubarov et al., Science **377** (6604), 425 (2022). Copyright 2022 Authors; licensed under a Creative Commons Attribution (CC BY) license.[72] (e) Band structure of time modulation in resonant materials. Reproduced with permission from Wang et al., Nat. Photonics **19** (2), 149-155 (2024). Copyright 2024 Authors; licensed under a Creative Commons Attribution (CC BY) license.[78] (f) Anomalous Maxwell-Garnett theory for PTCs. Reproduced with permission from Gong et al., Appl. Phys. Lett. **12** (3), 031414 (2025). Copyright 2025 Authors; licensed under a Creative Commons Attribution (CC BY) license.[60] (g) Free-electron radiation in PTCs. Reproduced with permission from Dikopoltsev et al., Proc. Natl. Acad. Sci. U.S.A.

119 (6), e2119705119 (2022). Copyright 2022 Authors; licensed under a Creative Commons Attribution (CC BY) license.[73] (h) Equivalent continuous model and discrete model, and topological edge states in the PTC with chiral symmetry. Reproduced with permission from Yang et al., ACS Photonics, 12 (5), 2389-2396(2025). Copyright 2025 Authors; licensed under a Creative Commons Attribution (CC BY) license.[65]

In addition, recent studies have shown that behaviors of free-electron radiation (such as Cherenkov and Smith-Purcell radiation) in PTCs are highly different from those in stationary materials.[68,73,81] For example, as shown in Fig. 4(g), Cherenkov radiation occurs in PTC even if the electron velocity is below the threshold (i.e., the phase velocity of light in host materials), which goes against the conventional wisdom. When associated with momentum-gap modes, the emission process of free electron is exponentially amplified, leading to significant quantum interference. Time modulation-induced free electron radiation not only significantly enhances intensity but also demonstrates strong robustness against variations in conventional structures. This robustness makes PTCs a powerful tool for stable and reconfigurable radiation generation, thus expanding their potential applications in high-precision optical control.

Recent studies have also extended the band topology from space photonic crystals to PTCs, giving birth to topological PTCs.[43,66,82–84] The advent of topological PTCs enables us to enhance the robustness of photonic modes in the time dimension, which is entirely a research gap in previous study of topological physics. For example, when the unit cell in time satisfies the time-inversion symmetry, each dispersion band of PTCs is associated with a quantized Zak phase.[43,85] These quantized Zak phases predict the emergence of time topological edge states localized at the time interface between two topologically distinct PTCs. Although the distinct gap characters determined by Zak phases guarantee the existence of time topological edge states, their eigenvalues (namely, eigen-wavevectors) tend to slightly oscillate as the disorder rate increases. To further enhance the robustness of topological time edge states, topological PTCs with chiral symmetry are proposed [Fig. 4(h)].[65] Condition of chiral symmetry is fulfilled if the ratio of time duration to the refractive index of each time slab is a constant value. The chiral symmetry quantizes the winding numbers of each dispersion band. The quantized winding number enables the prediction of chiral-symmetry protected time edge states. Provided the chiral symmetry is unbroken, the introduction of disorders and defects will never deviate the eigen-wavevectors of associated time edge states, exhibiting superior robustness.

**B. Spacetime modulated media**

**1. Spacetime interfaces and slabs**

A spacetime interface refers to an abrupt boundary where constitutive parameters move through space at a constant velocity. For example, $\varepsilon(x,t) = \varepsilon_1(x-vt<0)$, $\varepsilon(x,t) = \varepsilon_2(x-vt>0)$, where $v$ is the interface velocity. A spacetime slab is a finite modulated region comprising two or more such spacetime interfaces. Controlling reflection and transmission at spacetime interfaces is a fundamental research focus and an important engineering challenge for realizing dynamic and nonreciprocal photonic systems. Recently, the concept of quarter-wave impedance has been generalized into the spacetime domain.[52] This finding not only eliminates the reflection from both subluminal and superluminal interfaces, but also enables flexible frequency conversions via adjusting the interface velocity. To further enhance the controllability of the scattering process in the system of spacetime interface, the technique of coherent wave control is also generalized into the spacetime domain.[86] In the modulation of the spacetime interface, the identical magnitudes of frequency or wavevector are no longer a prerequisite to achieve interference between multiple incidences [Fig. 5(a)]. Destructive interference between the transmitted idler and reflected primary wave eliminates reflections, while transmission can also be suppressed by simply adjusting the incident wave amplitude. Moreover, the generalized coherent wave control enables the generation of high-frequency pulse through the interaction of low-frequency pulses with continuous waves. Recent work also reveals the spacetime diffraction by engineering the spacetime interface.[87] As illustrated in Fig. 5(b), by introducing ultrafast and high-contrast reflectance modulation in an indium tin oxide (ITO) thin film, a stationary interface can emulate a diffractive element moving at an arbitrary velocity. The diffracted light exhibits a linear correlation between its frequency and the outgoing angle, and the fringe spacing directly reflects the extent of separation in the spacetime double slit.

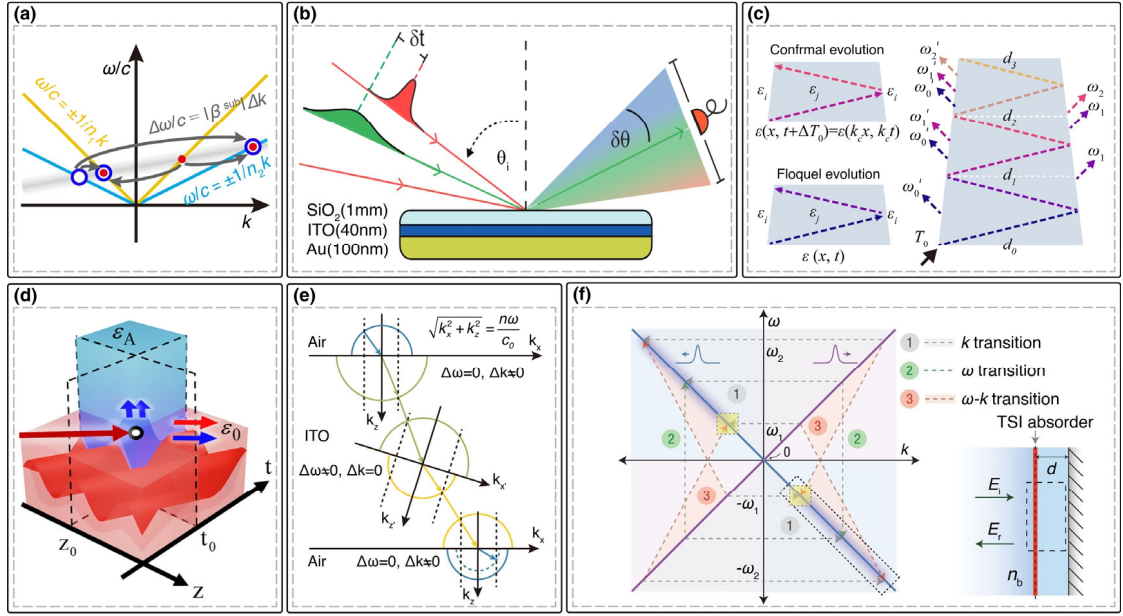

**FIG. 5.** Electromagnetic wave manipulation in spacetime modulations. (a) Schematic diagram of the principle of controlling generalized coherent waves. Reproduced with permission from Yu et al., Laser Photon. Rev. **19** (4), 2400399 (2024). Copyright 2024 Authors; licensed under a Creative Commons Attribution (CC BY) license.[86] (b) Schematic diagram of the spacetime diffraction. Reproduced with permission from Harwood et al., Nat. Commun. **16** (1), 5147 (2025). Copyright 2025 Authors; licensed under a Creative Commons Attribution (CC BY) license.[87] (c) Schematic of conformal spacetime modulation. Reproduced with permission from Wu et al., ACS Photonics **11** (8), 2992-3002 (2024). Copyright 2024 Authors; licensed under a Creative Commons Attribution (CC BY) license.[88] (d) Scattering behavior in the spacetime corner. Reproduced with permission from Stefanini et al., Phys. Rev. Lett. **135** (11), 113802 (2025). Copyright 2025 Authors; licensed under a Creative Commons Attribution (CC BY) license.[89] (e) Schematic diagram of beam deflection at a combined spacetime interface. Reproduced with permission from Fan et al., ACS Photonics **10** (8), 2467-2473 (2023). Copyright 2023 Authors; licensed under a Creative Commons Attribution (CC BY) license.[90] (f) Schematic of the principle of a TSI absorber. Reproduced with permission from Ciabattoni et al., Sci. Adv. **11** (3), eads7407 (2025). Copyright 2025 Authors; licensed under a Creative Commons Attribution (CC BY) license.[91]

A spacetime wedge forms if more than two spacetime interfaces with distinct velocities intersect with each other. Different from spatial wedges that produce diffraction, the spacetime wedges only support reflection and refraction due to causality.[92] In addition, cascaded Doppler effect induced by the spacetime wedges can lead to dynamic spectral broadening and reshaping. Through the cascaded Doppler shifts and energy injection, the spacetime wedges could also focus the free-space propagating waves, facilitating far-field super-resolution imaging and detection. Recent research also reports that the spacetime wedges enable geometric frequency combs.[88] To be specific, if the permittivity of the spacetime wedge behaves as a conformally evolving time-varying media [Fig. 5(c)], such structures are capable of generating geometric frequency combs with exponentially distributed spectra under monochromatic excitation. With its cascaded field enhancement and suppressed autocorrelation ambiguity, this geometric frequency comb is a promising light source for high-resolution radar and precision spectroscopy. Moreover, by folding aperiodic spacetime modulations, the spacetime Fresnel prism composed of multiple spacetime interfaces realizes compact anharmonic and nonreciprocal frequency conversion.[93]

Combining space and time interface gives rise to the formation of a spacetime corner. The interaction between waves and spacetime corner has demonstrated exotic scattering process, unlike those occurring at spatially homogeneous temporal interfaces. This distinction originates from the conflict between purely space and purely time boundary conditions, which is induced by the abrupt medium transition at the spacetime corner.[89] Specifically, as shown in Fig. 5(d), at $t=t_0$, time modulation is applied only to one half of the spatial domain, which introduces an electric field discontinuity at the interface $z=z_0$ for $t>t_0$. This conflict is reconciled by a localized equivalent magnetic surface current density that acts as a source, thereby generating shock waves. In addition, Combining space and time modulations can induce various types of combined spacetime interfaces, enabling versatile wave manipulation.[94] For instance, based on different modulation directions, perpendicular, parallel, or oblique combined spacetime interfaces can be formed, leading to phenomena such as temporal chirping, radiation emission, and temporal routing.

The combined spacetime interface also has a variety of electromagnetic wave modulation capabilities, such as wavefront shaping as shown in Fig. 5(e).[90] Propagation delay of the pump light induces a spatial refractive index gradient along its path, causing the optical beam to refract not only at the air-ITO interface but also within the ITO layer. Although the temporal refractive index gradient inside the ITO does not directly deflect the beam due to momentum conservation, frequency conversion at the air-ITO interface introduces a momentum offset, thereby altering the wave propagation direction. Since the deflection angle is also affected by the pump-signal angle, this platform can be used to achieve self-focusing effects. Another capability of the combined spacetime interface is to significantly enhance absorption.[91] To be specific, by enabling frequency coupling through time modulation, reflected components from different spectral channels are engineered to destructively interfere within the specific frequency bands, resulting in broadband reflection suppression and absorption performance. As marked by the black box in Fig. 5(f), the reflection at a specific frequency is the superposition of multiple frequency components, when the incident waves consist of two monochromatic components. Unlike conventional absorbers limited by the Rozanov limit,[95] the temporal-spatial interface absorber (TSI absorber) overcomes this constraint at sufficient modulation amplitude.

**2. Spacetime metamaterials**

Spacetime metamaterials refer to artificially engineered materials whose electromagnetic responses mainly arise from their subwavelength spatial ($a \ll \lambda$) and subperiod temporal ($T \ll 2\pi/\omega$) modulations. On a macroscopic scale, spacetime metamaterials can be treated as homogeneous effective media. Specifically, when the permittivity and permeability are modulated in the form of a traveling wave, their electromagnetic responses in the long-wavelength limit can be characterized by anisotropic parameters, which are equivalent to those of a moving matter system [Fig. 6(a)].[96] This equivalence has been rigorously established within the homogenization theory of spacetime metamaterials.[97] As illustrated in Fig. 6(c), the homogenization theory remains valid under both subluminal and superluminal modulations, and also applies to all frequencies where backward reflections are absent.

A key advantage of spacetime metamaterials is the ability to realize synthetic relativistic or even superluminal moving velocities, while achieving such high velocities within moving matter systems presents a formidable challenge and is sometimes impossible. For example, the traditional view holds that the realization of Fresnel drag relies on an actual matter motion, moreover, the effect becomes pronounced and experimentally measurable only when the motion velocity is sufficiently high (e.g., close to light speed in vacuum). Such a condition hinders the practical application of Fresnel drag. Recent work addresses this problem by proposing the platform of spacetime metamaterials to achieve the Fresnel drag.[96–98] As shown in Fig. 6(b), the Fresnel drag velocity opposes the modulation direction in the subluminal regime, while becoming consistent with the modulation direction in the superluminal regime.

Due to the broken time-reversal symmetry, spacetime metamaterials exhibit magnet-free nonreciprocity. As depicted in Fig. 6(d), a properly engineered modulation can induce nonreciprocal transitions between the free-space and evanescent waves, so called unidirectional evanescent wave conversion.[99] Furthermore, with judiciously modulation engineering, the spacetime metamaterials can also precisely control the amplitude and propagation direction of diffraction orders, thereby enabling operations such as frequency mixing and multiplication.[100] The spacetime metamaterials, with their unique physical mechanisms, hold great promise for magnetic-free circulators, isolators, nonreciprocal phase shifters and nonreciprocal upconversion focusing, etc.[101–103]

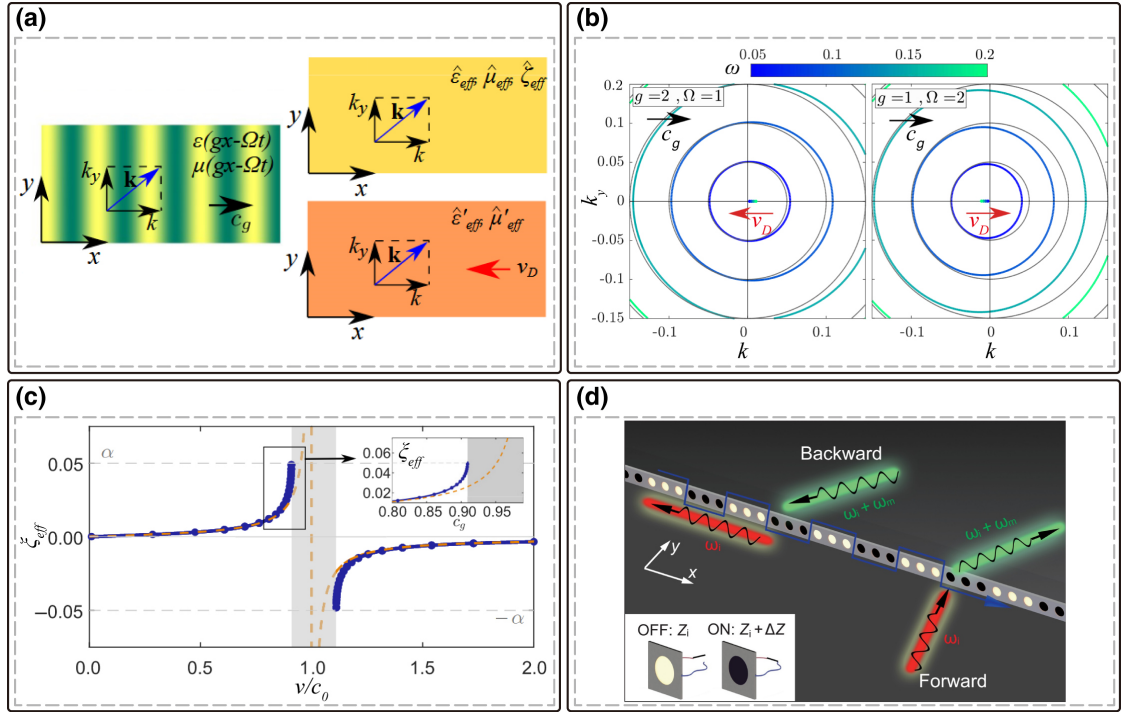

**FIG. 6.** Electromagnetic wave manipulation in spacetime metamaterials. (a) Equivalence of spacetime metamaterials with anisotropic medium and moving matter. (b) Fresnel drag in spacetime metamaterials, where the left and right sides correspond to the iso-frequency contours under subluminal and superluminal modulations. Reproduced with permission from Huidobro et al., Proc. Natl. Acad. Sci. U.S.A. **116** (50), 24943-24948 (2019). Copyright 2019 Authors; licensed under a Creative Commons Attribution (CC BY) license.[96] (c) The effective anisotropic coefficient $\xi_{eff}$ as a function of modulation velocity $v$. Reproduced with permission from Huidobro et al., Phys. Rev. Applied **16** (1), 014044 (2021). Copyright 2021 Authors; licensed under a Creative Commons Attribution (CC BY) license.[97] (d) Nonreciprocal effect in spacetime metamaterials. Reproduced with permission from Chen et al., Sci. Adv. **7** (45), eabj1198 (2021). Copyright 2021 Authors; licensed under a Creative Commons Attribution (CC BY) license.[99]

### 3. Spacetime photonic crystals

Spacetime photonic crystals (STPCs) are realized when the constitutive parameters of a medium are periodically modulated in both space and time (e. g., $\varepsilon(\boldsymbol{r}+\boldsymbol{a},t+T)=\varepsilon(\boldsymbol{r},t)$). STPCs are essentially a dynamic extension of traditional photonic crystals in the time dimension. Although both spacetime metamaterials and spacetime photonic crystals are spatiotemporal composite materials, they differ fundamentally in both concept and physics. Spacetime metamaterials, as exotic effective homogeneous media constructed from subwavelength and subperiod spacetime modulation units, exhibit physical properties determined by the filling factor of these modulation units. In contrast, spacetime photonic crystals are wavelength-scale spacetime periodic structures whose behavior is dictated by the Floquet-Bloch theorem. They are primarily used to generate dynamic bandgaps and to manipulate the energy and momentum of light. Traveling-wave modulated STPCs can exhibit distinct bandgaps.[31,70] In the subluminal regime, energy bandgaps open, resembling those of conventional space photonic crystals [Figs. 7(a-i) and 7(a-ii)], whereas in the superluminal regime, momentum bandgaps emerge, analogous to PTCs [Figs. 7(a-iii) and 7(a-iv)]. The energy bandgaps exhibit exponential spatial decay, characterized by real frequencies and complex wavevectors, whereas the momentum bandgaps feature exponential temporal growth, characterized by complex frequencies and real wavevectors. The band structure of STPCs is highly dependent on the spacetime modulation of permittivity and permeability. Under different modulations, the band structures can be exploited for various applications. For example, when only the permittivity is modulated, the asymmetric characteristic of the bandgap leads to non-degenerate splitting between forward- and backward-propagating modes at the Brillouin zone, enabling magnet-free optical isolation.[104] If the permittivity is spatially aperiodic but temporally periodic, and the permeability is spatially aperiodic, selectively enhancing or suppressing specific frequencies can be achieved.[105] It provides a theoretical basis for the design of spatiotemporal mixers. Furthermore, if the modulation depths of permittivity and permeability are balanced, the originally tilted spatiotemporal bandgap can fully close.[106] This balanced modulation not only enhances nonreciprocity but also enables perfect reflectionless transmission. Overall, the diverse modulation strategies in STPCs provide new possibilities for developing high-performance reconfigurable microwave photonic devices.[107,108,99,109]

STPCs with traveling-wave modulation in the interluminal regime are particularly intriguing, as they give rise to a wealth of distinctive phenomena. Recent studies have shown that when the modulation follows a sinusoidal traveling wave, broadband dispersion degeneracy emerges within the interluminal regime.[110] This distinctive dispersion enables intra-band coupling among all forward-propagating waves, thereby facilitating broadband nonreciprocal amplification of incident signals [Fig. 7(b)], including even DC components.[70] This absence of bandgaps also permits simultaneous phase-coherent growth of multiple harmonics, which can be utilized for designing cascaded parametric oscillators.[111] Moreover, interluminal STPCs host a series of singularities that either concentrate or disperse energy, akin to event horizons of optical white or black holes.[112] At these singular points, interluminal modes can also induce stimulated emission of photon pairs, providing a potential platform for observing Hawking radiation in laboratory.

In addition to frequency and momentum bandgaps, STPCs with standing-wave modulation can also give rise to a unique mixed energy-momentum bandgap, representing an intermediate regime that combines features of both.[113] A prominent feature of the mixed energy-momentum bandgap is the simultaneous complexity of both frequency and wavevector [Fig. 7(c-i)], where their propagation dynamics are determined by the dominant mechanism. When the imaginary frequency component exceeds the imaginary wavevector component [Fig. 7(c-ii)], propagation characteristics align with that of energy bandgap modes, exhibiting spatial exponential decay. On the contrary, the behavior follows that of momentum bandgap modes, manifesting amplitude exponential growth [Fig. 7(c-iii)]. More surprisingly, when the imaginary parts of both are equal, the mixed energy-momentum bandgap collapses into a single point, simplifying its spacetime band structure to a continuous straight line. In this case, the incident wave preserves its original propagation direction and phase velocity, while simultaneously exciting a counter-propagating component. Energy and momentum increase linearly in time and evolve in opposite directions, analogous to the mechanism of electromagnetic radiation.

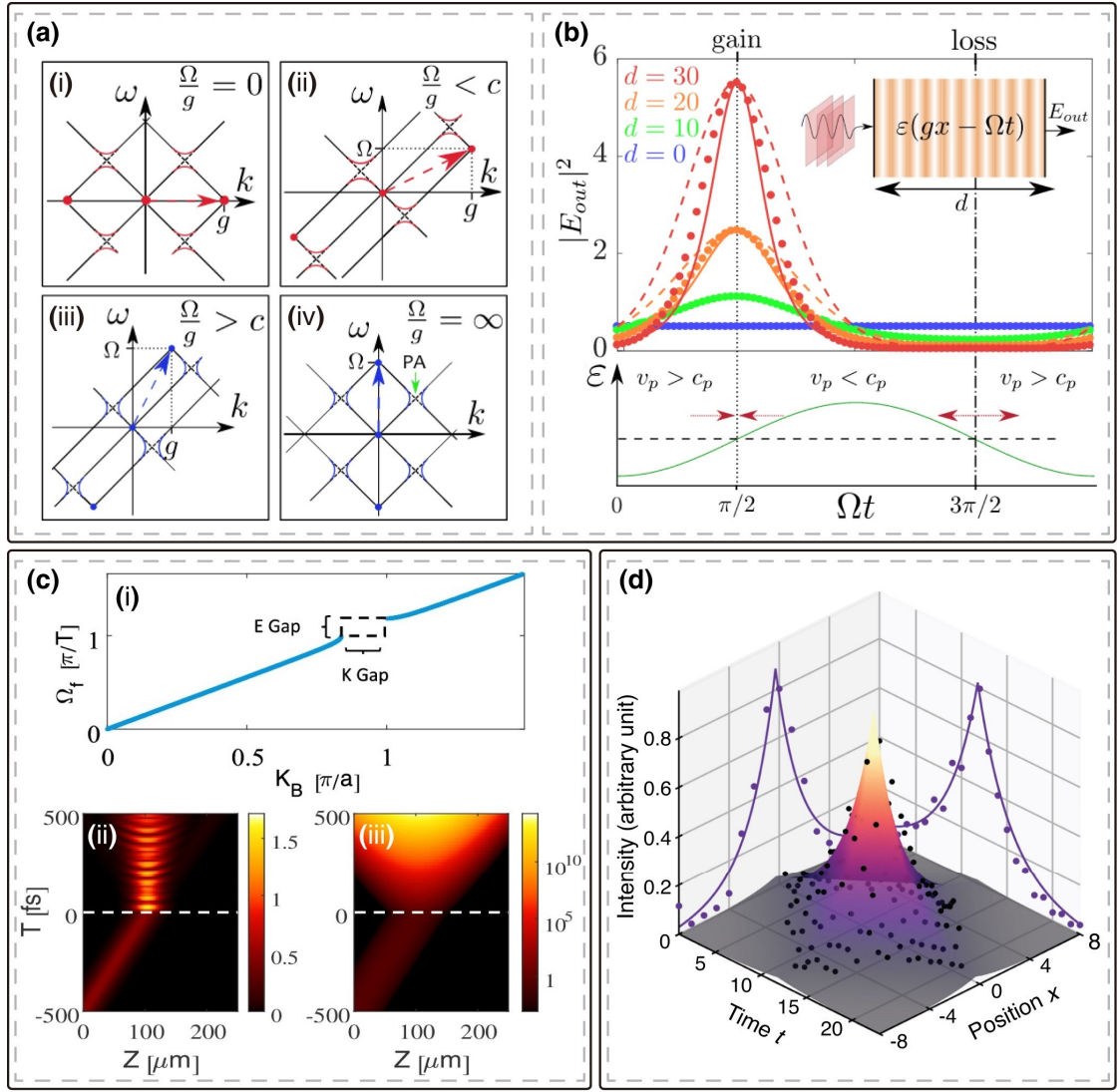

FIG. 7. Electromagnetic wave manipulation in STPCs. (a) Band structures for SM (i), subluminal STM (ii), superluminal STM (iii), and TM (iv) cases. (b) Transmission characteristics in a luminal spacetime metamaterial. Reproduced with permission from Galiffi et al., Phys. Rev. Lett. **123** (20), 206101 (2019). Copyright 2019 Authors; licensed under a Creative Commons Attribution (CC BY) license.[70] (c) The mixed energy-momentum bandgap in STPCs (i) and corresponding space-time field distribution (ii, iii). Reproduced with permission from Sharabi et al., Optica **9** (6), 585 (2022). Copyright 2022 Authors; licensed under a Creative Commons Attribution (CC BY) license.[113] (d) Spacetime topological event in STCPs. Reproduced with permission from Joshua et al., Nat. Photonics. **19** (5), 518-525 (2025). Copyright 2025 Authors; licensed under a Creative Commons Attribution (CC BY) license.[114]

The amplification in STPCs is unstable but can be controlled by topologically induced spacetime-localized states. For instance, spacetime topological events can be observed by using a discrete-time quantum walk experimental platform.[114] At the intersection of spacetime domain walls, as shown in Fig. 7(d), spacetime topological events emerge with exponentially localized along both the space and time axis. The events exhibit unique causality-suppressed coupling effects, where excitation can only occur through the past light cone, with coupling from the future light cone being strictly suppressed. Furthermore, when disorder perturbations are introduced into the system, the spacetime topological events show the selective collapse characteristics. Specifically, under strong disorder conditions, spatial localization vanishes while time localization is retained. This finding provides a controllable quantum simulation platform for studying spacetime topology.

Spacetime topological events can also manifest in both nonlinear and linear systems.[115,116] In nonlinear systems, when the nonlinear effects balance spacetime amplification and attenuation, soliton solutions can arise that simultaneously exhibit both spatial and temporal localization. In linear systems, by extending the classical Jackiw-Rebbi model to the spacetime dimensions, topological event wave packets with exponential decay characteristics are obtained at the domain wall intersections. Furthermore, periodically arranging domain walls in spacetime forms topological event lattices, which suppress noise amplification and significantly enhance system stability. These studies together represent a significant leap in advancing spacetime topology from theoretical concepts to experimental validation. As research progresses, STPCs shows greater potential for applications in areas, such as ultrafast light manipulation, topological lasers, and novel communication systems.[72,117,118]

## IV. EXPERIMENTAL IMPLEMENTATION AND VALIDATION

Owing to the continuous development of artificial materials (e.g., nonlinear materials and metasurfaces) and the maturation of ultrafast modulation technologies, experimental demonstrations on time-varying media have expanded from the microwave to the optical regime, exhibiting an impressive capability for broadband and cross-spectral implementation. Various physical platforms, including transmission lines,[119–124] nonlinear optical media,[45,125,126] metasurfaces,[127–129] and time-synthetic lattices,[82,130,131] have been employed to implement time-varying media and to experimentally verify a wide range of phenomena arising from time or spacetime modulation.

Transmission line platforms are among the earliest and most widely adopted approaches, owing to structural simplicity, high precision control, and compatibility with traditional electronic circuits. A typical configuration incorporates periodically loaded tunable elements, such as electronically triggered capacitor arrays, optically activated photodiodes, or voltage-controlled varactors. As a signal propagates, the switches can be synchronized to toggle at a prescribed instant, abruptly changing the line's effective impedance and forming a temporal interface. For example, Fig. 8(a) illustrates microstrip lines that employ high-speed photodiodes, triggered via a fiber network, to introduce additional capacitance,[123] which reduces the switching time to the picosecond range, about an order of magnitude faster than nanosecond-level electronic switches. Such platforms have enabled observations of temporal reflection, phase conjugation, time reversal, frequency conversion, momentum bandgaps, and coherent control.[119–124] However, the reliance on discrete components constrains scalability at higher frequencies, where device response times and insertion loss become significant limiting factors. These limitations have motivated the development of optical platforms that support ultrafast, broadband time modulation.

In the optical regime, nonlinear effects provide a natural route to implementing time-varying media. Mechanisms such as the Kerr effect,[132–135] free-carrier dispersion,[136,137] laser-induced plasma,[138,139] and optomechanical interactions[140] can induce significant refractive index changes on femtosecond to picosecond timescales. ENZ materials, such as ITO, are particularly promising, offering large permittivity tunability up to 0.7 near the zero-permittivity point with minimal absorption.[141] As shown in Fig. 8(b), illuminating an ITO film with a pump beam and adjusting the pump-probe delay can control second harmonic generation.[125] Such platforms show outstanding flexibility and bandwidth advantages, and have achieved phenomena such as temporal diffraction (single-slit and double-slit analogies) and

red- or blue-shifts.[52,77] However, achieving large modulation depths in the optical regime often requires high pump powers, which increases both energy consumption and the risk of material damage. These challenges have prompted the search for compact, integrable platforms with embedded active control.

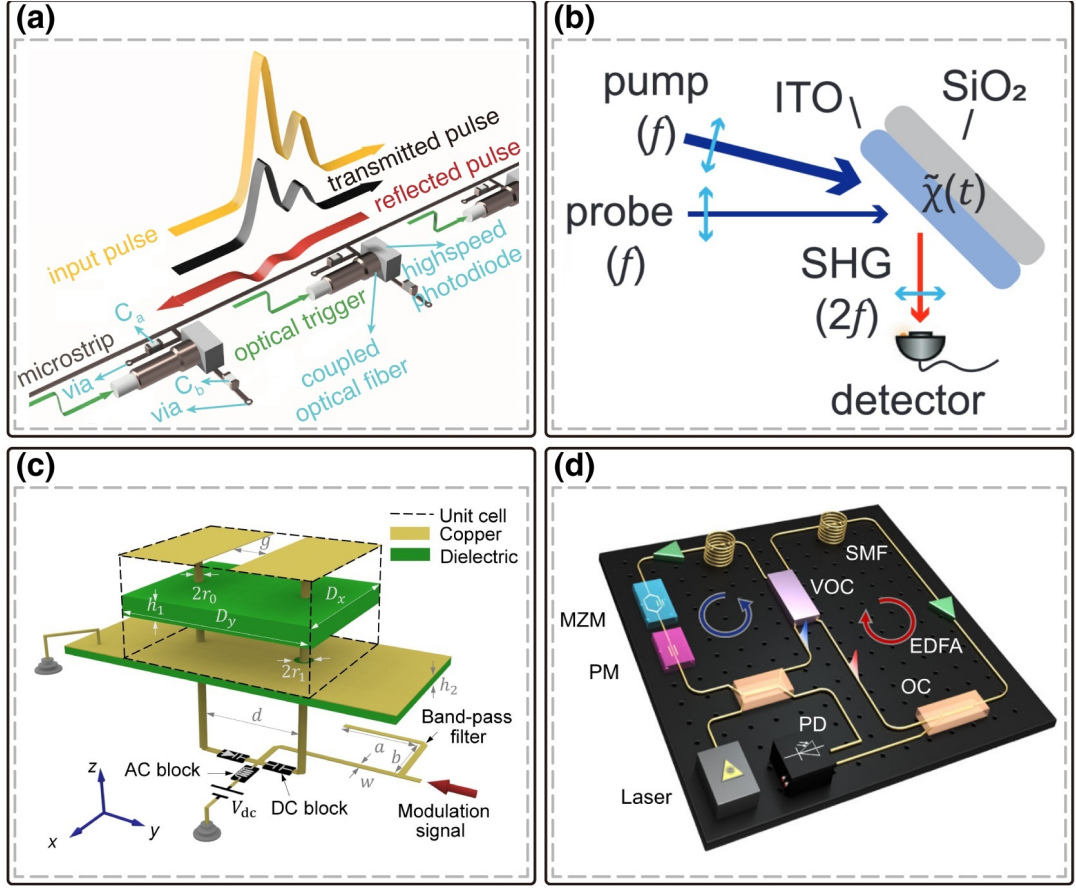

**FIG. 8.** Representative platforms for realizing time-varying media. (a) High-speed photodiode-based microstrip transmission line: optical triggering via a fiber network induces abrupt capacitance changes, forming temporal interface. Reproduced with permission from Jones et al., Nat. Commun. **15** (1), 6786 (2024). Copyright 2024 Authors; licensed under a Creative Commons Attribution (CC BY) license.[123] (b) Nonlinear media platform based on indium tin oxide (ITO): pump-light excitation produces ultrafast refractive-index jumps to create temporal interface. Reproduced with permission from Tirole et al., Nat. Commun. **15** (1), 7752 (2024). Copyright 2024 Authors; licensed under a Creative Commons Attribution (CC BY) license.[125] (c) Varactor-loaded metasurface: tuning the DC bias modulates the unit-cell capacitance, enabling spatiotemporal control. Reproduced with permission from Ye et al., Laser Photon. Rev. **19** (10), 2401345 (2025). Copyright 2025 Authors; licensed under a Creative Commons Attribution (CC BY) license.[127] (d) Time-synthetic lattice with dual coupled fiber loops: the relative position of pulse maps to spatial position $n$, while pulse round trips map to discrete time steps $m$. The temporal interface is introduced by adjusting variable optical couplers (OC), Mach-Zehnder modulators (MAM), and phase modulators (PM). Reproduced with permission from Ren et al., Nat. Commun. **16** (1), 707 (2025). Copyright 2025 Authors; licensed under a Creative Commons Attribution (CC BY) license.[82]

Metasurfaces represent a key direction toward this goal. Metasurfaces, including those with electrically controlled plasmas or varactor-loaded unit cells, can undergo abrupt impedance changes on sub-nanosecond timescales, enabling spatiotemporal control of electromagnetic fields. As depicted in Fig. 8(c), their hardware generally consists of a dielectric substrate, periodic metallic patterns with active tuning components, and control circuitry comprising microstrip splitters and DC bias networks.[127] By tuning the effective capacitance, metasurfaces have been used to observe momentum bandgap, parametric amplification, mixed energy-momentum bandgap closure, manipulate harmonics, and reveal modulation depth driven topological transitions (e.g., capacitive-to-inductive phase switching).[127–129] While metasurfaces offer ultrafast control, strong field localization, and compatibility with planar integration, their tuning range is often limited by available capacitance swing, and stable operation at higher frequencies remains technically challenging. These constraints have encouraged exploration of more flexible synthetic dimensions.

Time-synthetic lattices discretize temporal evolution into a programmable lattice. As shown in Fig. 8(d), a dual-coupled fiber-loop architecture maps the relative position of pulse to a discrete spatial position $n$, while the number of pulse round trips encodes discrete time steps $m$.[82] Tunable optical couplers (OC), Mach-Zehnder modulators (MAM), and phase modulators (PM) provide independent control of coupling and phase, enabling programmable evolution in the lattice. Using this framework, experiments have demonstrated temporal scattering, momentum bandgaps, temporal topological edge states, and collapse of super-Bloch oscillations.[82,130,131] Their primary strengths lie in high parameter tunability, precise control, and the ability to simulate diverse physical systems. However, the high experimental complexity and limited integration with practical devices constrain their near-term engineering applicability.

As discussed above, each platform offers unique advantages. Transmission lines excel in low-frequency demonstrations and fundamental proof of principle studies. Nonlinear optical media enable ultrafast responses and broad spectral tunability. Metasurfaces combine speed with integration potential. Time-synthetic lattices excel at exploring novel physical mechanisms and programmable dynamics Future progress will require simultaneous improvements in modulation velocity and depth, reduction of power consumption, and simplification of system architectures. Promising directions include the development of new materials with large nonlinear coefficients, low loss, and broadband response, as well as architectures that integrate programmable photonics with reconfigurable electromagnetic platforms. These advances could enable practical applications of time-varying media in on-chip and wireless communications, signal processing, quantum information, and topological photonics.

**V. CONCLUSION AND OUTLOOK**

This review has systematically categorized time-varying media by their modulation strategies and provided a comprehensive overview of their underlying physical mechanisms, structural platforms, and application prospects. By actively modulating constitutive parameters in the temporal domain, time-varying media break the intrinsic constraints of continuously temporal translation symmetry and open a new paradigm for dynamic wave manipulation. We clarified the principles and distinct electromagnetic characteristics of time modulations and spacetime modulations, highlighting their distinctive response phenomena. On the application front, we discussed opportunities in frequency conversion, nonreciprocal transport, signal shaping, broadband control, parametric amplification, and topological photonics. For experimental comparability, we summarize four predominant platforms for realizing time-varying media and compared their relative merits. Despite substantial progress, as shown in Fig. 9, several directions remain especially promising for further breakthroughs.

In rarely explored spacetime interface, one particularly promising direction is the light-matter interaction in the interluminal regime [Fig. 9(a)]. As mentioned previously, materials with interluminal modulation exhibit remarkable scattering phenomena. Most of these effects remain unverified experimentally, and some are not fully resolved theoretically. One example is the emergence of broadband dispersion degeneracy under sinusoidal traveling wave modulations,[70] which has been predicted to enable broadband nonreciprocal amplification and to produce Hawking-like radiation in correlated photon pairs.[142,112] However, these unique phenomena are solely predicted in theory, awaiting experimental observation. The effect would be more striking under the circumstances of interluminal step modulation, where a single dynamic interface leads to single- or triple-wave scattering as mentioned in the section of principles of time-varying media. Such exotic scattering behaviors naturally induce the following paradox: two spacetime boundary conditions overdetermine one scattering wave and underdetermine three scattering waves, such that conventional approaches fail to solve these two fundamental problems. In 1975, L. A. Ostrovski proposed an analytical treatment by virtually inserting a gradient-index slab at the interluminal interface to balance boundary conditions and obtain the three amplitudes.[53] Nevertheless, the validity of this approach remains controversial due to the absence of experimental confirmation. The lack of verification primarily stems from the stringent requirements for realizing step-modulated interluminal interfaces, which demand precise control of modulation velocity and ultrafast material response times (significantly shorter than the temporal period of the studied waves). Meanwhile, the overdetermined case still lacks a closed-form analytical solution. Both theory and experiment are therefore urgently needed to uncover the scattering mechanisms at interluminal interfaces.

As a more general class, accelerated-velocity spacetime modulations remain largely unexplored [Fig. 9(a)]. Very few recent studies predict that interactions with accelerated interfaces can induce asymmetric spectral broadening, dynamic Fabry-Pérot resonances, non-uniform Doppler shifts, gravitational-like bending, and two-photon emission.[143–147] As with interluminal modulation, these results are presently theoretical and unverified. On the theory side, there is a clear opportunity to generalize effective-medium and Bloch formulations to accelerated backgrounds, ideally within a covariant framework that treats acceleration explicitly and enforces causality, passivity, and energy conservation. Within such a framework, acceleration becomes an independent design degree of freedom for spacetime metamaterials and STPCs, enabling dispersion engineering and stability analysis beyond constant-velocity spacetime modulations. Experimentally, candidate routes include accelerating refractive-index fronts via chirped pump pulses in ENZ/nonlinear films, spacetime engineered electro-optic modulators with time-varying drive curvature, and moving-plasma/photocarrier interfaces in semiconductors, each route trades off achievable acceleration, loss, and damage thresholds. These directions would further enhance controllability of light-matter interactions and provide testable benchmarks for accelerated-velocity spacetime modulation theory.

Beyond the well-established traveling-wave modulations, standing-wave and traveling-standing-wave modulations in periodic modulation time-varying media remain rarely explored [Fig. 9(b)]. Periodically modulated time-varying media provide a powerful route to break temporal translation symmetry and engineer wave dynamics, enabling unique control over frequency, momentum, and energy exchange. To date, most studies have focused on traveling-wave modulations, which induce nonreciprocal responses and have been extensively investigated both theoretically and experimentally.[31] These efforts have established a mature framework for nonreciprocal devices, unidirectional transparency, parametric amplification, and topological photonics.[32,107] From a physical perspective, traveling-wave modulation effectively emulates an "artificial moving medium" or a "moving photonic crystal", introducing direction-dependent coupling in the Floquet band structure and thereby enabling intrinsic asymmetry between forward and backward propagation. In contrast, standing-wave modulation does not impart an effective momentum bias, preserving symmetric coupling between positive- and negative-frequency sidebands. This intrinsic symmetry makes standing-wave modulation particularly suitable for time-domain signal processing, including spectral reshaping, temporal lensing, and pulse manipulation, and establishes it as a fundamental platform for dynamic spectral engineering. Despite this promise, research on standing-wave modulation remains in its early stages. In particular, topological phenomena in such systems have only recently begun to attract attention.[116] This imbalance underscores the need for systematic theoretical and experimental investigations. In this broader context, traveling- and standing-wave modulations represent two complementary extremes, associated respectively with directional asymmetry and spectral symmetry. While their individual advantages are now increasingly clear, the superposition of these two forms, namely, traveling-standing wave modulation, remains largely unexplored. The absence of a unified framework for such hybrid modulation schemes has significantly constrained the functional integration of time-varying platforms. Looking ahead, time-varying media incorporating traveling-standing-wave modulation offer a promising pathway toward simultaneously achieving nonreciprocal transport and symmetric spectral engineering within a single architecture. When combined with tailored spacetime symmetry design, such systems may further host topologically protected temporal states, enabling the coexistence of nonreciprocity, spectral tunability, and topological robustness in a unified platform. Beyond wave-based functionalities, periodically modulated structures also provide a fertile ground for studying radiation phenomena arising from the interaction between charged particles and time-varying media. In particular, free-electron radiation such as Cherenkov radiation, Smith-Purcell radiation, and high-order processes induced by free electrons traversing periodically modulated media offers rich opportunities for tunable radiation generation and spatiotemporal control.[148–150] These platforms enable precise engineering of the spectral, angular, and temporal characteristics of electron-induced radiation, opening new avenues for compact radiation sources and advanced spatiotemporal radiation manipulation.

Random modulation, particularly random spacetime modulation, has been scarcely explored in time-varying media [fig. 9(c)]. In space modulated system, random modulation, such as spatial disorders, leads to Anderson localization, where destructive interference results in exponential localization of wave functions in space.[151] Similarly, recent studies on time modulation have shown that temporal disorders can also profoundly affect wave propagation.[61–63] For instance, in disordered PTCs, the group velocity of a pulse can decrease exponentially and eventually approach zero, while its intensity grows exponentially. The statistics follow the single-parameter scaling characteristic of Anderson localization. Weak temporal disorders can also map a three-dimensional medium onto an Anderson model, inducing Anderson localization in the time domain. A natural direction for future research is to investigate disorders in STPCs. In such systems, disorders may induce localization with coupled space-time characteristics and reshape the band structure. In addition, disorder-driven closing and reopening of bandgaps may modify the associated topological invariants in STPCs, potentially triggering topological phase transitions. These effects raise the intriguing possibility of realizing spatiotemporal analogues of Anderson topological insulators, supporting topologically protected modes in the combined space and time domain. Investigating these regimes may uncover new connections between disorders, topology, and spacetime dynamics in time-varying media.

The integration of time-varying media with quantum theory will provide a more comprehensive description of the relevant optical processes. In this regime, temporal variation and quantum fluctuations are intrinsically linked, leading to a variety of nonclassical phenomena that extend far beyond conventional photonics [Fig. 9(d)]. For example, quantum fluctuations at temporal interfaces can induce the Casimir effect, generating photon pairs directly from vacuum.[152–154] Time-varying media can further amplify vacuum fluctuations while controlling emission angles and frequencies.[155] In addition, intriguing effects include frequency shifts in quantum antireflection coatings, Hawking-like quantum radiation in synthetic moving gratings, and photon emission arising from accelerated refractive-index modulation.[156,112,157,145,158] Exploring and expanding these phenomena opens an emerging research avenue that bridges quantum field theory and time-varying media. Leveraging the quantum nature of time-varying media could enable the development of tailored quantum light sources, controllable quantum state manipulation schemes, photonic platforms for simulating relativistic quantum effects, and so on.[159,160] In the long term, this direction may provide powerful tools for advancing quantum communication, computation, and precision sensing technologies, thereby enriching the broader landscape of quantum photonics.

| directions \ dimensions | | (e) 1+1D theory | (e) 1+1D exp. | (e) 2+1D theory | (e) 2+1D exp. | (e) 3+1D theory | (e) 3+1D exp. |
|---|---|---|---|---|---|---|---|
| (a) rarely explored spacetime interface | interluminal interface | 53, 70, 112, 142 | | | | | |
| | accelerated interface | 143, 145, 147 | | 144, 146 | | | |
| (b) rarely explored periodic modulation | standing-wave modulation | 116 | | | | | |
| | traveling-standing-wave modulation | | | | | | |
| (c) unexplored modulation | random modulation | | | | | | |
| (d) microscopic description | quantum mechanics | 112, 157, 159 | | 155 | | | |

▨ no theory/experiment

**FIG. 9** Outlook of time-varying media: (a) Interluminal interface and accelerated interface in rarely explored spacetime interface. (b) Standing-wave modulation and traveling-standing-wave modulation in rarely explored periodic modulation. (c) Random modulation in unexplored modulation. (d) Integration of time-varying media with quantum mechanics in microscopic description. (e) Dimensions of time-varying media.

While time modulation has revealed rich physics, broader impact will come from higher dimensional control that unifies spatial and temporal manipulation [Fig. 9(e)]. Most current studies are largely confined to one-dimensional (1D) geometries due to the inherent complexities of material nonlinear response and theoretical modeling. Future endeavors could strive to break these spatial constraints by integrating 2D metasurfaces or 3D topological structures with dynamic modulation to construct multidimensional time-varying photonic crystals. Such holistic control would not only enable a leap in parametric manipulation, such as facilitating arbitrary polarization evolution, complex wavefront reconfiguration, and orbital angular momentum steering, but also gives rise to novel physical effects unattainable in lower-dimensional systems, such as nonreciprocal photonic topological states and spacetime exceptional points (EPs).[78,161] Realizing this goal requires the synergistic advancement of material science and integration technology, where the development of ultrafast response, low-loss, and high modulation depth materials is coupled with scalable photonic architectures. These advancements will not only deepen our fundamental understanding of time-varying media but also accelerate the transition of time-varying electromagnetics into practical, high-impact applications.

In summary, the future of time-varying media research is expected to develop along the directions of deeper theoretical foundations, higher-dimensional architectures, and stronger interdisciplinary integration. With the further refinement of the theoretical framework, innovation in quantum control mechanisms, and continuous exploration of advanced technologies, time-varying media hold promise as transformative enablers for next-generation communication, sensing, energy, computation, and imaging technologies.

## ACKNOWLEDGMENTS

This work was supported by National Natural Science Foundation of China (Grant No. 12404363, 62422106, 92573203), Natural Science Foundation of Jiangsu Province (Grant No. BK20241374), Distinguished Professor Fund of Jiangsu Province, Fundamental Research Funds for the Central Universities, NUAA (Grants No. NS2024022, NE2024007), and the Shanghai Pujiang Program (Grant No. 23PJ1411500).

## AUTHOR DECLARATIONS

**Conflict of Interest**

The authors have no conflicts to disclose.

**Author Contributions**

Youxiu Yu and Hao Hu contributed equally to this paper.

**Youxiu Yu**: Data curation (lead); Writing-original draft (lead); Writing-review & editing (equal). **Hao Hu:** Funding acquisition (lead); Supervision (equal); Writing-review & editing (equal). **Qianru Yang:** Writing-review & editing (equal). **Linyang Zou:** Writing-review & editing (equal). **Dongjue Liu:** Writing-review & editing (equal); Funding acquisition (supporting). **Hao Chi Zhang:** Writing-review & editing (equal); Funding acquisition (supporting). **Yu Luo:** Funding acquisition (lead); Project administration (lead); Supervision (lead).

## DATA AVAILABILITY

The data that supports the findings of this study are available from the corresponding authors upon reasonable request.

## REFERENCES

[1] Kong, J. A, *Electromagnetic Wave Theory* (EMW Publishing, Cambridge Massachusetts, 2008).

[2] Sengupta, D. L. and Sarkar, T. K., "Maxwell, Hertz, the Maxwellians, and the early history of electromagnetic waves," IEEE Antennas Propag. Mag. **45**(2), 13–19 (2003).

[3] C.M. Soukoulis, and M. Wegener, "Past achievements and future challenges in the development of three-dimensional photonic metamaterials," Nat. Photonics **5**(9), 523–530 (2011).

[4] C. Wang, X. Chen, Z. Gong, R. Chen, H. Hu, H. Wang, Y. Yang, L. Tony, B. Zhang, H. Chen, and X. Lin, "Superscattering of light: fundamentals and applications," Rep. Prog. Phys. **87**(12), 126401 (2024).

[5] D.R. Smith, J.B. Pendry, and M.C.K. Wiltshire, "Metamaterials and negative refractive index," Science **305**(5685), 788–792 (2004).

[6] J. Pendry, L. Sw, A. Holden, W. Stewart, I. Youngs, and D.H. Heath, "Extremely low frequency plasmons in metallic microstructures," Phys. Rev. Lett. **76**(25), 4773 (1996).

[7] Viktor G Veselago, "The electrodynamics of substances with simultaneously negative values of ε and μ," Sov. Phys. Usp. **10**(4), 509 (1968).

[8] D. R. Smith, Willie J. Padilla, D. C. Vier, S. C. Nemat-Nasser, and S. Schultz, "Composite medium with simultaneously negative permeability and permittivity," Phys. Rev. Lett. **84**(18), 4184–4187 (2000).

[9] Shelby, R. A., Smith, D. R., and Schultz, S., "Experimental verification of a negative index of refraction," Science **292**(5514), 77–79 (2001).

[10] Smolyaninov, Igor I., Hung, Yu-Ju, and Davis, Christopher C., "Magnifying superlens in the visible frequency range," Science **315**(5819), 1699–1701 (2007).

[11] Watts, Claire M., Liu, Xianliang, and Padilla, Willie J., "Metamaterial electromagnetic wave absorbers,"

Adv. Mater. **24**(23), OP98–OP120 (2012).

[12] Filippo Capolino, *Theory and Phenomena of Metamaterials* (CRC Press, Boca Raton, 2017).

[13] J.B. Pendry, D. Schurig, and D.R. Smith, "Controlling electromagnetic fields," Science **312**(5781), 1780–1782 (2006).

[14] J. Zhang, J.B. Pendry, and Y. Luo, "Transformation optics from macroscopic to nanoscale regimes: a review," Adv. Photon. **1**(01), 1 (2019).

[15] D. Schurig, J.J. Mock, B.J. Justice, S.A. Cummer, J.B. Pendry, A.F. Starr, and D.R. Smith, "Metamaterial electromagnetic cloak at microwave frequencies," Science **314**(5801), 977–980 (2006).

[16] E. Yablonovitch, "Inhibited spontaneous emission in solid-state physics and electronics," Phys. Rev. Lett. **58**(20), 2059–2062 (1987).

[17] S. John, "Strong localization of photons in certain disordered dielectric superlattices," Phys. Rev. Lett. **58**(23), 2486–2489 (1987).

[18] J. D. Joannopoulos, S. G. Johnson, J. N. Winn, and R. D. Meade, *Photonic Crystals: Molding the Flow of Light*, Second Edition (Princeton University Press, 2008).

[19] J. D. Joannopoulos, Pierre R. Villeneuve, and Shanhui Fan, "Photonic crystals: putting a new twist on light," Nature **386**(6621), 143–149 (1997).

[20] P. Russell, "Photonic crystal fibers," Science **299**(5605), 358–362 (2003).

[21] Noda, Susumu, Fujita, Masayuki, and Asano, Takashi, "Spontaneous-emission control by photonic crystals and nanocavities," Nat. Photonics **1**(8), 449–458 (2007).

[22] R.V. Nair, and R. Vijaya, "Photonic crystal sensors: An overview," Prog. Quantum Electron. **34**(3), 89–134 (2010).

[23] F.D.M. Haldane, and S. Raghu, "Possible realization of directional optical waveguides in photonic crystals with broken time-reversal symmetry," Phys. Rev. Lett. **100**(1), 013904 (2008).

[24] Z. Wang, Y.D. Chong, J.D. Joannopoulos, and M. Soljačić, "Reflection-free one-way edge modes in a gyromagnetic photonic crystal," Phys. Rev. Lett. **100**(1), 013905 (2008).

[25] K. Kawabata, K. Shiozaki, M. Ueda, and M. Sato, "Symmetry and topology in non-Hermitian physics," Phys. Rev. X **9**(4), 041015 (2019).

[26] M.C. Rechtsman, J.M. Zeuner, Y. Plotnik, Y. Lumer, D. Podolsky, F. Dreisow, S. Nolte, M. Segev, and A. Szameit, "Photonic Floquet topological insulators," Nature **496**(7444), 196–200 (2013).

[27] Y. Li, M. Jung, Y. Yu, Y. Han, B. Zhang, and G. Shvets, "Topological directional coupler," Laser Photon. Rev. **18**(11), 2301313 (2024).

[28] R. Duggan, S.A. Mann, and A. Alù, "Nonreciprocal photonic topological order driven by uniform optical pumping," Phys. Rev. B **102**(10), 100303 (2020).

[29] Y.-G. Sang, J.-Y. Lu, Y.-H. Ouyang, H.-Y. Luan, J.-H. Wu, J.-Y. Li, and R.-M. Ma, "Topological polarization singular lasing with highly efficient radiation channel," Nat. Commun. **13**(1), 6485 (2022).

[30] A. Hashemi, M.J. Zakeri, P.S. Jung, and A. Blanco-Redondo, "Topological quantum photonics," APL Photonics **10**(1), 010903 (2025).

[31] M.M. Asgari, P. Garg, X. Wang, M.S. Mirmoosa, C. Rockstuhl, and V. Asadchy, "Theory and applications of photonic time crystals: a tutorial," Adv. Opt. Photonics **16**(4), 958 (2024).

[32] E. Galiffi, R. Tirole, S. Yin, H. Li, S. Vezzoli, P.A. Huidobro, M.G. Silveirinha, R. Sapienza, A. Alù, and J.B. Pendry, "Photonics of time-varying media," Adv. Photonics **4**(01), 014002 (2022).

[33] S. Yin, E. Galiffi, and A. Alù, "Floquet metamaterials," eLight **2**(1), 8 (2022).

[34] N. Engheta, "Four-dimensional optics using time-varying metamaterials," Science **379**(6638), 1190–1191 (2023).

[35] C. Caloz, A. Bahrami, and A. Stevens, "Structuring space-time for photon and electron waves: opinion," Opt. Mater. Express **15**(4), 711 (2025).

[36] F. Bloch, "Nuclear induction," Phys. Rev. **70**(7–8), 460–474 (1946).

[37] F.R. Morgenthaler, "Velocity modulation of electromagnetic waves," IEEE Trans. Microwave Theory Techn. **6**(2), 167–172 (1958).

[38] Oliner A A and Hessel A, "Wave propagation in a medium with a progressive sinusoidal disturbance," IRE Trans. Microw. Theory Tech. **9**(4), 337–343 (1961).

[39] C. Caloz, and Z.-L. Deck-Leger, "Spacetime metamaterials-part I: general concepts," IEEE Trans. Antennas Propag. **68**(3), 1569–1582 (2020).

[40] C. Caloz, and Z.-L. Deck-Leger, "Spacetime metamaterials-part II: theory and applications," IEEE Trans. Antennas Propag. **68**(3), 1583–1598 (2020).

[41] C. Jung, E. Lee, and J. Rho, "The rise of electrically tunable metasurfaces," Sci. Adv. **10**(34), eado8964 (2024).

[42] L. Zhang, and T.J. Cui, "Space-time-coding digital metasurfaces: principles and applications," Research **2021**, 9802673 (2021).

[43] E. Lustig, Y. Sharabi, and M. Segev, "Topological aspects of photonic time crystals," Optica **5**(11), 1390 (2018).

[44] L.A. Hall, M. Yessenov, S.A. Ponomarenko, and A.F. Abouraddy, "The space-time Talbot effect," APL Photonics **6**(5), 056105 (2021).

[45] Y. Zhou, M.Z. Alam, M. Karimi, J. Upham, O. Reshef, C. Liu, A.E. Willner, and R.W. Boyd, "Broadband frequency translation through time refraction in an epsilon-near-zero material," Nat. Commun. **11**(1), 2180 (2020).

[46] V. Bacot, M. Labousse, A. Eddi, M. Fink, and E. Fort, "Time reversal and holography with spacetime transformations," Nat. Phys. **12**(10), 972–977 (2016).

[47] D.L. Sounas, and A. Alù, "Non-reciprocal photonics based on time modulation," Nat. Photonics **11**(12), 774–783 (2017).

[48] H. Li, S. Yin, and A. Alù, "Nonreciprocity and Faraday rotation at time interfaces," Phys. Rev. Lett. **128**(17), 173901 (2022).

[49] Jacob B. Khurgin, "Photonic time crystals and parametric amplification: similarity and distinction," ACS Photonics **11**(6), 2150–2159 (2024).

[50] I. Liberal, A. Ganfornina-Andrades, and J.E. Vázquez-Lozano, "Spatiotemporal symmetries and energy-momentum conservation in uniform spacetime metamaterials," ACS Photonics **11**(12), 5273–5280 (2024).

[51] Z.-L. Deck-Léger, N. Chamanara, M. Skorobogatiy, M.G. Silveirinha, and C. Caloz, "Uniform-velocity spacetime crystals," Adv. Photonics **1**(05), 056002 (2019).

[52] Y. Yu, H. Hu, L. Zou, Q. Yang, X. Lin, Z. Li, L. Gao, and D. Gao, "Antireflection spatiotemporal metamaterials," Laser Photon. Rev. **17**(9), 2300130 (2023).

[53] L.A. Ostrovskil, "Some 'moving boundaries paradoxes' in electrodynamics," Sov. Phys. Usp. **18**(6), 452–458 (1975).

[54] A. Ortega-Gomez, M. Lobet, J.E. Vázquez-Lozano, and I. Liberal, "Tutorial on the conservation of momentum in photonic time-varying media [Invited]," Opt. Mater. Express **13**(6), 1598 (2023).

[55] C. Caloz, Z.-L. Deck-Léger, A. Bahrami, O.C. Vicente, and Z. Li, "Generalized space-time engineered modulation (GSTEM) metamaterials: a global and extended perspective," IEEE Antennas Propag. Mag. **65**(4), 50–60 (2023).

[56] R. Tirole, S. Vezzoli, E. Galiffi, I. Robertson, D. Maurice, B. Tilmann, S.A. Maier, J.B. Pendry, and R. Sapienza, "Double-slit time diffraction at optical frequencies," Nat. Phys. **19**(7), 999–1002 (2023).

[57] V. Pacheco-Peña, and N. Engheta, "Antireflection temporal coatings," Optica **7**(4), 323 (2020).

[58] V. Pacheco-Peña, and N. Engheta, "Temporal equivalent of the Brewster angle," Phys. Rev. B **104**(21), 214308 (2021).

[59] V. Pacheco-Peña, and N. Engheta, "Effective medium concept in temporal metamaterials," Nanophotonics **9**(2), 379–391 (2020).

[60] Z. Gong, R. Chen, H. Chen, and X. Lin, "Anomalous Maxwell-Garnett theory for photonic time crystals," Appl. Phys. Lett. **12**(3), 031414 (2025).

[61] R. Pierrat, J. Rocha, and R. Carminati, "Causality and instability in wave propagation in random time-varying media," Phys. Rev. Lett. **134**(23), 233801 (2025).

[62] Y. Sharabi, E. Lustig, and M. Segev, "Disordered photonic time crystals," Phys. Rev. Lett. **126**(16), 163902 (2021).

[63] K.S. Eswaran, A.E. Kopaei, and K. Sacha, "Anderson localization in photonic time crystals," Phys. Rev. B **111**(18), L180201 (2025).

[64] M.R. Shcherbakov, K. Werner, Z. Fan, N. Talisa, E. Chowdhury, and G. Shvets, "Photon acceleration and tunable broadband harmonics generation in nonlinear time-dependent metasurfaces," Nat. Commun. **10**(1), 1345 (2019).

[65] Y. Yang, H. Hu, L. Liu, Y. Yang, Y. Yu, Y. Long, X. Zheng, Y. Luo, Z. Li, and F.J. Garcia-Vidal, "Topologically protected edge states in time photonic crystals with chiral symmetry," ACS Photonics **12**(5), 2389–2396 (2025).

[66] M.-W. Li, J.-W. Liu, X. Wang, W.-J. Chen, and G. Ma, "Topological temporal boundary states in a non-Hermitian spatial crystal," Phys. Rev. Lett. **135**(18), 187101 (2025).

[67] X. Ni, S. Yin, H. Li, and A. Alù, "Topological wave phenomena in photonic time quasicrystals," Phys. Rev. B **111**(12), 125421 (2025).

[68] J.-F. Zhu, A. Nussupbekov, Y. Fan, W. Zhou, Z. Song, X. Wang, Z.-W. Zhang, C.-H. Du, X. Wei, P. Bai, C.E. Png, C.-W. Qiu, and L. Wu, "Smith-Purcell radiation from time grating," Newton **1**(2), 100023 (2025).

[69] V. Pacheco-Peña, and N. Engheta, "Temporal aiming," Light Sci. Appl. **9**(1), 129 (2020).

[70] E. Galiffi, P.A. Huidobro, and J.B. Pendry, "Broadband nonreciprocal amplification in luminal metamaterials," Phys. Rev. Lett. **123**(20), 206101 (2019).

[71] Akbarzadeh, A, Chamanara, N, and Caloz, C, "Inverse prism based on temporal discontinuity and spatial dispersion," Opt. Lett. **43**(14), 3297–3300 (2018).

[72] M. Lyubarov, Y. Lumer, A. Dikopoltsev, E. Lustig, Y. Sharabi, and M. Segev, "Amplified emission and lasing in photonic time crystals," Science **377**(6604), 425–428 (2022).

[73] A. Dikopoltsev, Y. Sharabi, M. Lyubarov, Y. Lumer, S. Tsesses, E. Lustig, I. Kaminer, and M. Segev, "Light emission by free electrons in photonic time-crystals," Proc. Natl. Acad. Sci. U.S.A. **119**(6), e2119705119 (2022).

[74] J. Park, K. Lee, R.-Y. Zhang, H.-C. Park, J.-W. Ryu, G.Y. Cho, M.Y. Lee, Z. Zhang, N. Park, W. Jeon, J. Shin, C.T. Chan, and B. Min, "Spontaneous emission decay and excitation in photonic time crystals," Phys. Rev. Lett. **135**(13), 133801 (2025).

[75] C. Rizza, G. Castaldi, and V. Galdi, "Spin-controlled photonics via temporal anisotropy," Nanophotonics **12**(14), 2891–2904 (2023).

[76] H. Li, S. Yin, H. He, J. Xu, A. Alù, and B. Shapiro, "Stationary charge radiation in anisotropic photonic

time crystals," Phys. Rev. Lett. **130**(9), 093803 (2023).

[77] Y. Hu, H. Hao, J. Zhang, M. Tong, X. Cheng, and T. Jiang, "Anisotropic temporal metasurfaces for tunable ultrafast photoactive switching dynamics," Laser Photon. Rev. **15**(10), 2100244 (2021).

[78] X. Wang, P. Garg, M.S. Mirmoosa, A.G. Lamprianidis, C. Rockstuhl, and V.S. Asadchy, "Expanding momentum bandgaps in photonic time crystals through resonances," Nat. Photonics **19**(2), 149–155 (2024).

[79] S. Zhang, J. Dong, H. Li, J. Xu, and B. Shapiro, "Longitudinal optical phonons in photonic time crystals containing a stationary charge," Phys. Rev. B **110**(10), L100306 (2024).

[80] J. Feinberg, D.E. Fernandes, B. Shapiro, and M.G. Silveirinha, "Plasmonic Time Crystals," Phys. Rev. Lett. **134**(18), 183801 (2025).

[81] X. Gao, X. Zhao, X. Ma, and T. Dong, "Free electron emission in vacuum assisted by photonic time crystals," J. Phys. D: Appl. Phys. **57**(31), 315112 (2024).

[82] Y. Ren, K. Ye, Q. Chen, F. Chen, L. Zhang, Y. Pan, W. Li, X. Li, L. Zhang, H. Chen, and Y. Yang, "Observation of momentum-gap topology of light at temporal interfaces in a time-synthetic lattice," Nat. Commun. **16**(1), 707 (2025).

[83] T. Sheppard, C.B. Camacho, S. Weidemann, A. Szameit, J. Feis, F. Schindler, and H.M. Price, "Topological localisation in time from PT symmetry," arXiv Preprint **2509.06679**, (2025).

[84] S. Tong, Q. Zhang, G. Li, K. Zhang, C. Xie, and C. Qiu, "Observation of momentum-band topology in PT-symmetric acoustic Floquet lattices," Nat. Commun. **16**, 9975 (2025).

[85] Lin, Mi, Ahmed, Shakeel, Jamil, Maryam, Liang, Zixian, Wang, Qiong, and Ouyang, Zhengbiao, "Temporally-topological defect modes in photonic time crystals," Opt. Express **32**(6), 9820–9836 (2024).

[86] Y. Yu, D. Gao, Y. Yang, L. Liu, Z. Li, Q. Yang, H. Wu, L. Zou, X. Lin, J. Xiong, S. Hou, L. Gao, and H. Hu, "Generalized coherent wave control at dynamic interfaces," Laser Photon. Rev. **19**(4), 2400399 (2024).

[87] A.C. Harwood, S. Vezzoli, T.V. Raziman, C. Hooper, R. Tirole, F. Wu, S.A. Maier, J.B. Pendry, S.A.R. Horsley, and R. Sapienza, "Space-time optical diffraction from synthetic motion," Nat. Commun. **16**(1), 5147 (2025).

[88] H. Wu, Q. Yang, H. Hu, L. Zou, X. Wang, J. He, S. Pan, Y. Zheng, T.J. Cui, and Y. Luo, "Conformal spatiotemporal modulation enabled geometric frequency combs," ACS Photonics **11**(8), 2992–3002 (2024).

[89] L. Stefanini, E. Galiffi, S. Yin, S. Singh, D.M. Solís, N. Engheta, A. Toscano, D. Ramaccia, F. Bilotti, and A. Alù, "Theory and experimental observation of scattering by a space-time corner," Phys. Rev. Lett. **135**(11), 113802 (2025).

[90] Q. Fan, A.M. Shaltout, J. Van De Groep, M.L. Brongersma, and A.M. Lindenberg, "Ultrafast wavefront shaping via space-time refraction," ACS Photonics **10**(8), 2467–2473 (2023).

[91] M. Ciabattoni, Z. Hayran, and F. Monticone, "Observation of broadband super-absorption of electromagnetic waves through space-time symmetry breaking," Sci. Adv. **11**(3), eads7407 (2025).

[92] K.D. Kinder, A. Bahrami, and C. Caloz, "Doppler pulse amplification," Phys. Rev. Applied **24**(4), 044029 (2025).

[93] Z. Li, X. Ma, A. Bahrami, Z.-L. Deck-Léger, and C. Caloz, "Space-time Fresnel prism," Phys. Rev. Applied **20**(5), 054029 (2023).

[94] V. Pacheco-Peña, M. Fink, and N. Engheta, "Temporal chirp, temporal lensing and temporal routing via space-time interfaces," Phys. Rev. B **111**(10), L100306 (2025).

[95] K.N. Rozanov, "Ultimate thickness to bandwidth ratio of radar absorbers," IEEE Trans. Antennas

Propag. **48**(8), 1230–1234 (2000).

[96] P.A. Huidobro, E. Galiffi, S. Guenneau, R.V. Craster, and J.B. Pendry, "Fresnel drag in space–time-modulated metamaterials," Proc. Natl. Acad. Sci. U.S.A. **116**(50), 24943–24948 (2019).

[97] P.A. Huidobro, M.G. Silveirinha, E. Galiffi, and J.B. Pendry, "Homogenization theory of space-time metamaterials," Phys. Rev. Applied **16**(1), 014044 (2021).

[98] C. Caloz, A. Alù, S. Tretyakov, D. Sounas, K. Achouri, and Z.-L. Deck-Léger, "Electromagnetic nonreciprocity," Phys. Rev. Applied **10**(4), 047001 (2018).

[99] Z. Chen, Y. Peng, H. Li, J. Liu, Y. Ding, B. Liang, X.-F. Zhu, Y. Lu, J. Cheng, and A. Alù, "Efficient nonreciprocal mode transitions in spatiotemporally modulated acoustic metamaterials," Sci. Adv. **7**(45), eabj1198 (2021).

[100] S. Moreno-Rodríguez, A. Alex-Amor, P. Padilla, J.F. Valenzuela-Valdés, and C. Molero, "Space-time metallic metasurfaces for frequency conversion and beamforming," Phys. Rev. Applied **21**(6), 064018 (2024).

[101] X. Wang, A. Díaz-Rubio, H. Li, S.A. Tretyakov, and A. Alù, "Theory and design of multifunctional space-time metasurfaces," Phys. Rev. Applied **13**(4), 044040 (2020).

[102] Y. Hadad, D.L. Sounas, and A. Alu, "Space-time gradient metasurfaces," Phys. Rev. B **92**(10), 100304 (2015).

[103] S. Taravati, and G.V. Eleftheriades, "Microwave space-time-modulated metasurfaces," ACS Photonics **9**(2), 305–318 (2022).

[104] N. Chamanara, S. Taravati, Z.-L. Deck-Léger, and C. Caloz, "Optical isolation based on space-time engineered asymmetric photonic band gaps," Phys. Rev. B **96**(15), 155409 (2017).

[105] S. Taravati, "Aperiodic space-time modulation for pure frequency mixing," Phys. Rev. B **97**(11), 115131 (2018).

[106] Sajjad Taravati, "Giant linear nonreciprocity, zero reflection, and zero band gap in equilibrated space-time-varying media," Phys. Rev. Applied **9**, 064012 (2018).

[107] Z. Yu, and S. Fan, "Complete optical isolation created by indirect interband photonic transitions," Nat. Photonics **3**(2), 91–94 (2009).

[108] S. Taravati, N. Chamanara, and C. Caloz, "Nonreciprocal electromagnetic scattering from a periodically space-time modulated slab and application to a quasisonic isolator," Phys. Rev. B **96**(16), 165144 (2017).

[109] M.H. Mostafa, M.S. Mirmoosa, E. Galiffi, S. Yin, A. Alù, and S.A. Tretyakov, "Broadband amplification of light through adiabatic spatiotemporal modulation," arXiv Preprint **2506.20358**, (2025).

[110] E. Galiffi, M.G. Silveirinha, P.A. Huidobro, and J.B. Pendry, "Photon localization and Bloch symmetry breaking in luminal gratings," Phys. Rev. B **104**(1), 014302 (2021).

[111] Q. Yang, H. Hu, X. Li, and Y. Luo, "Cascaded parametric amplification based on spatiotemporal modulations," Photon. Res. **11**(5), B125 (2023).

[112] S.A.R. Horsley, and J.B. Pendry, "Quantum electrodynamics of time-varying gratings," Proc. Natl. Acad. Sci. U.S.A. **120**(36), e2302652120 (2023).

[113] Y. Sharabi, A. Dikopoltsev, E. Lustig, Y. Lumer, and M. Segev, "Spatiotemporal photonic crystals," Optica **9**(6), 585 (2022).

[114] J. Feis, S. Weidemann, T. Sheppard, H.M. Price, and A. Szameit, "Space-time-topological events in photonic quantum walks," Nat. Photonics **19**(5), 518–525 (2025).

[115] L. Zhang, Z. Fan, and Y. Pan, "Event soliton formation in mixed energy-momentum gaps of nonlinear spacetime crystals," arXiv Preprint **2503.16113**, (2025).

116 L. Zhang, Z. Zhao, Q. Pan, C. Pan, Q. Cheng, and Y. Pan, "Generating and weaving topological event wavepackets in photonic spacetime crystals with fully energy-momentum gapped," arXiv Preprint **2507.15309**, (2025).

117 F. Biancalana, A. Amann, A.V. Uskov, and E.P. O'Reilly, "Dynamics of light propagation in spatiotemporal dielectric structures," Phys. Rev. E **75**(4), 046607 (2007).

118 A.M. Shaltout, V.M. Shalaev, and M.L. Brongersma, "Spatiotemporal light control with active metasurfaces," Science **364**(6441), eaat3100 (2019).

119 J.R. Reyes-Ayona, and P. Halevi, "Observation of genuine wave vector (*k* or *β*) gap in a dynamic transmission line and temporal photonic crystals," Appl. Phys. Lett. **107**(7), 074101 (2015).

120 E. Galiffi, D.M. Solís, S. Yin, N. Engheta, and A. Alù, "Electrodynamics of photonic temporal interfaces," Light Sci. Appl. **14**, 338 (2025).

121 H. Moussa, G. Xu, S. Yin, E. Galiffi, Y. Ra'di, and A. Alù, "Observation of temporal reflection and broadband frequency translation at photonic time interfaces," Nat. Phys. **19**(6), 863–868 (2023).

122 E. Galiffi, G. Xu, S. Yin, H. Moussa, Y. Ra'di, and A. Alù, "Broadband coherent wave control through photonic collisions at time interfaces," Nat. Phys. **19**(11), 1703–1708 (2023).

123 T.R. Jones, A.V. Kildishev, M. Segev, and D. Peroulis, "Time-reflection of microwaves by a fast optically-controlled time-boundary," Nat. Commun. **15**(1), 6786 (2024).

124 J.R. Reyes-Ayona, and P. Halevi, "Electromagnetic wave propagation in an externally modulated low-pass transmission line," IEEE Trans. Microw. Theory Tech. **64**(11), 3449–3459 (2016).

125 R. Tirole, S. Vezzoli, D. Saxena, S. Yang, T.V. Raziman, E. Galiffi, S.A. Maier, J.B. Pendry, and R. Sapienza, "Second harmonic generation at a time-varying interface," Nat. Commun. **15**(1), 7752 (2024).

126 A.Yu. Bykov, J. Deng, G. Li, and A.V. Zayats, "Time-dependent ultrafast quadratic nonlinearity in an epsilon-near-zero platform," Nano Lett. **24**(12), 3744–3749 (2024).

127 X. Wang, M.S. Mirmoosa, V.S. Asadchy, C. Rockstuhl, S. Fan, and S.A. Tretyakov, "Metasurface-based realization of photonic time crystals," Sci. Adv. **9**(14), eadg7541 (2023).

128 J. Sisler, Prachi Thureja, Meir Y. Grajower, Ruzan Sokhoyan, Ivy Huang, and Harry A. Atwater, "Electrically tunable space-time metasurfaces at optical frequencies," Nat. Nanotechnol. **19**, 1491–1498 (2024).

129 X. Ye, Y.G. Wang, J.F. Yao, Y. Wang, C.X. Yuan, and Z.X. Zhou, "Realization of spatiotemporal photonic crystals based on active metasurface," Laser Photon. Rev. **19**(10), 2401345 (2025).

130 X. Hu, S. Wang, C. Qin, C. Liu, L. Zhao, Y. Li, H. Ye, W. Liu, S. Longhi, P. Lu, and B. Wang, "Observing the collapse of super-Bloch oscillations in strong-driving photonic temporal lattices," Adv. Photonics **6**(04), (2024).

131 C. Qin, H. Ye, S. Wang, L. Zhao, M. Liu, Y. Li, X. Hu, C. Liu, B. Wang, S. Longhi, and P. Lu, "Observation of discrete-light temporal refraction by moving potentials with broken Galilean invariance," Nat. Commun. **15**(1), 5444 (2024).

132 Dekker, R, Driessen, A, Wahlbrink, T, Moormann, C, Niehusmann, J, and Foerst, M, "Ultrafast Kerr-induced all-optical wavelength conversion in silicon waveguides using 1.55 μm femtosecond pulses," Opt. Express **14**(18), 8336–8346 (2006).

133 J.B. Khurgin, M. Clerici, and N. Kinsey, "Fast and slow nonlinearities in epsilon-near-zero materials," Laser Photon. Rev. **15**(2), (2021).

134 M. Zahirul Alam, Israel De Leon, and Robert W. Boyd, "Large optical nonlinearity of indium tin oxide in its epsilon-near-zero region," Science **352**(6287), 795–797 (2016).

135 J. Bohn, T.S. Luk, C. Tollerton, S.W. Hutchings, I. Brener, S. Horsley, W.L. Barnes, and E. Hendry,

"All-optical switching of an epsilon-near-zero plasmon resonance in indium tin oxide," Nat. Commun. **12**(1), (2021).

[136] Stefan F. Preble, Qianfan Xu, and Michal Lipson, "Changing the colour of light in a silicon resonator," Nat. Photonics **1**(5), 293–296 (2007).

[137] K. Kondo, and T. Baba, "Dynamic wavelength conversion in copropagating slow-light pulses," Phys. Rev. Lett. **112**(22), 223904 (2014).

[138] Yablonovitch, E., "Self-phase modulation of light in a laser-breakdown plasma," Phys. Rev. Lett. **32**(20), 1101–1104 (1974).

[139] N.C. Lopes, G. Figueira, J.M. Dias, L.O. Silva, J.T. Mendonça, P. Balcou, G. Rey, and C. Stenz, "Laser pulse frequency up-shifts by relativistic ionization fronts," Europhys. Lett. **66**(3), 371–377 (2004).

[140] L. Fan, C.-L. Zou, M. Poot, R. Cheng, X. Guo, X. Han, and H.X. Tang, "Integrated optomechanical single-photon frequency shifter," Nat. Photonics **10**(12), 766–770 (2016).

[141] J. Bohn, T.S. Luk, S. Horsley, and E. Hendry, "Spatiotemporal refraction of light in an epsilon-near-zero indium tin oxide layer: frequency shifting effects arising from interfaces," Optica **8**(12), 1532 (2021).

[142] Pendry, J. B., Galiffi, E., and Huidobro, P. A., "Photon conservation in trans-luminal metamaterials," Optica **9**(7), 724–730 (2022).

[143] K.D. Kinder, A. Bahrami, and C. Caloz, "Scattering and chirping at accelerated interfaces," arXiv Preprint **2506.19575**, (2025).

[144] A. Bahrami, Z.-L. Deck-Léger, and C. Caloz, "Electrodynamics of accelerated-modulation space-time metamaterials," Phys. Rev. Applied **19**(5), 054044 (2023).

[145] J. Sloan, N. Rivera, J.D. Joannopoulos, and M. Soljačić, "Controlling two-photon emission from superluminal and accelerating index perturbations," Nat. Phys. **18**(1), 67–74 (2022).

[146] A. Bahrami, Z.-L. Deck-Léger, Z. Li, and C. Caloz, "A generalized FDTD scheme for moving electromagnetic structures with arbitrary space–time configurations," IEEE Trans. Antennas Propag. **72**(2), 1721–1734 (2024).

[147] A. Bahrami, K. De Kinder, and C. Caloz, "Arbitrary pulse shaping using accelerated interfaces," Small Struct., e202500340 (2025).

[148] Z. Gong, J. Chen, R. Chen, X. Zhu, C. Wang, X. Zhang, H. Hu, Y. Yang, B. Zhang, H. Chen, I. Kaminer, and X. Lin, "Interfacial Cherenkov radiation from ultralow-energy electrons," Proc. Natl. Acad. Sci. U.S.A. **120**(38), e2306601120 (2023).

[149] Z. Gong, R. Chen, Z. Wang, X. Xi, Y. Yang, B. Zhang, H. Chen, I. Kaminer, and X. Lin, "Free-electron resonance transition radiation via Brewster randomness," Proc. Natl. Acad. Sci. U.S.A. **122**(6), e2413336122 (2025).

[150] Bowen Zhang, Zheng Gong, Ruoxi Chen, Xuhuinan Chen, Yi Yang, Hongsheng Chen, Ido Kaminer, and Xiao Lin, "Reversed Cherenkov radiation via Fizeau-Fresnel drag," Appl. Phys. Rev. **12**(4), 041421 (2025).

[151] M. Segev, Y. Silberberg, and D.N. Christodoulides, "Anderson localization of light," Nat. Photonics **7**(3), 197–204 (2013).

[152] J. Sloan, N. Rivera, J.D. Joannopoulos, and M. Solja, "Casimir light in dispersive nanophotonics," Phys. Rev. Lett. **127**(5), 053603 (2021).

[153] S.A. Hassani Gangaraj, G.W. Hanson, and F. Monticone, "Dynamical Casimir effects: The need for nonlocality in time-varying dispersive nanophotonics," Phys. Rev. A **110**(4), L041502 (2024).

[154] J.E. Sustaeta-Osuna, F.J. García-Vidal, and P.A. Huidobro, "Quantum theory of photon pair creation in photonic time crystals," ACS Photonics **12**(4), 1873–1880 (2025).

[155] J.E. Vázquez-Lozano, and I. Liberal, "Shaping the quantum vacuum with anisotropic temporal boundaries," Nanophotonics **12**(3), 539–548 (2023).

[156] I. Liberal, J.E. Vázquez-Lozano, and V. Pacheco-Peña, "Quantum antireflection temporal coatings: quantum state frequency shifting and inhibited thermal noise amplification," Laser & Photonics Reviews **17**(9), 2200720 (2023).

[157] S.A.R. Horsley, and J.B. Pendry, "Travelling wave amplification in stationary gratings," Phys. Rev. Lett. **133**(15), 156903 (2024).

[158] Z. Dong, H. Li, T. Wan, Q. Liang, Z. Yang, and B. Yan, "Quantum time reflection and refraction of ultracold atoms," Nat. Photon. **18**(1), 68–73 (2024).

[159] J.B. Pendry, and S.A.R. Horsley, "QED in space–time varying materials," APL Quantum **1**(2), 020901 (2024).

[160] M.S. Mirmoosa, T. Setälä, and A. Norrman, "Quantum state engineering and photon statistics at electromagnetic time interfaces," Phys. Rev. Research **7**(1), 013120 (2025).

[161] P. Garg, E. Almpanis, L. Zimmer, J.D. Fischbach, X. Wang, M.S. Mirmoosa, M. Nyman, N. Stefanou, N. Papanikolaou, V. Asadchy, and C. Rockstuhl, "Photonic time crystals assisted by quasi-bound states in the continuum," arXiv Preprint **2507.15644**, (2025).